\documentstyle[aps,preprint,eqsecnum,epsfig]{revtex}
\tightenlines
\begin{document}
\preprint{UFIFT-HEP-98-41}
\date{\today}

\title{The caustic ring singularity}
\author{P. Sikivie}
\address{Department of Physics, University of Florida, Gainesville, FL 
32611}

\maketitle

\begin{abstract}
I investigate the caustics produced by the fall of collisionless dark
matter in and out of a galaxy in the limit of negligible velocity 
dispersion.  The outer caustics are spherical shells enveloping the 
galaxy.  The inner caustics are rings.  These are located near where 
the particles with the most angular momentum are at their distance of 
closest approach to the galactic center.  The surface of a caustic ring 
is a closed tube whose cross-section is a $D_{-4}$ catastrophe.  It has 
three cusps amongst which exists a discrete $Z_3$ symmetry.  A detailed 
analysis is given in the limit where the flow of particles is axially 
and reflection symmetric and where the transverse dimensions of the 
ring are small compared to the ring radius.  Five parameters describe 
the caustic in that limit.  The relations between these parameters and 
the initial velocity distribution of the particles are derived.  The 
structure of the caustic ring is used to predict the shape of the bump 
produced in a galactic rotation curve by a caustic ring lying in the 
galactic plane.     
\end{abstract}

\pacs{PACS numbers: xyz}

\narrowtext

\section{Introduction}

There are compelling reasons to believe that the dark matter of the 
universe is constituted, at least in part, by non-baryonic collisionless 
particles with low primordial velocity dispersion \cite{kt}.  Such
particles are called cold dark matter.  The leading candidates are axions 
and weakly interacting massive particles (WIMPs).  Before the onset of 
galaxy formation but after the time $t_{eq}$ of equality between matter
and radiation, the velocity dispersion of the cold dark matter candidates 
is very small, of order $\delta v_a (t) \sim 3\cdot 10^{-17} 
\left( {10^{-5} eV\over m_a}\right)~\left({t_0\over t}\right)^{2/3}$ for 
axions and $\delta v_W (t) \sim 10^{-11} 
\left({GeV\over m_W}\right)^{1/2}~\left({t_0\over t}\right)^{2/3}$ for WIMPs, 
where $t_0$ is the present age of the universe and $m_a$ and $m_W$ are 
respectively the masses of the axion and the WIMP.  The above estimates of 
the primordial velocity dispersions, $\delta v_a$ and $\delta_W$, are very 
crude but our point is only that in the context of this paper, and of galaxy 
formation in general, $\delta v_a$ and $\delta v_w$ are entirely negligible.  
Massive neutrinos, on the other hand, have primordial velocity dispersion 
$\delta v_\nu (t) \simeq 5.3~10^{-4} \left({eV\over 
m_\nu}\right)~\left({t_0\over t}\right)^{2/3}$ which is comparable to the 
virial velocity in galaxies and is therefore non-negligible in the context 
of galaxy formation \cite{STG}.  For this reason, massive neutrinos are 
called ``hot dark matter''.

Before the onset of galaxy formation, the collisionless dark matter 
particles lie on a thin 3-dimensional (3D) sheet in 6D phase-space.  The 
thickness of this sheet is the primordial velocity dispersion $\delta v$.
If each of the aforementioned species of collisionless particles is present, 
the phase-space sheet has three layers, a very thin layer of axions, a medium 
layer of WIMPs and a thick layer of neutrinos.  The phase-space 
sheet is located on the 3D hypersurface of points $(\vec{r}, 
\vec{v})~:~\vec{v} = H(t)\vec{r} + \Delta \vec{v} (\vec{r},t)$ where $H(t) 
= {2\over 3t}$ is the Hubble expansion rate and $\Delta \vec{v} 
(\vec{r},t)$ is the peculiar velocity field.  Fig. 1 shows a 2D section of 
6D phase-space along the $(z,\dot{z})$ plane.  The wiggly line is the 
intersection of the 3D sheet on which the particles lie in phase-space 
with the plane of the figure.  The thickness of the line is the velocity 
dispersion $\delta v$, whereas the amplitude of the wiggles in the line 
is the peculiar velocity $\Delta v$.  If there were no peculiar 
velocities, the line would be straight since $\dot{z} = H(t) z$ in that case.

The peculiar velocities are associated with density perturbations and grow 
by gravitational instability as $\Delta v\sim t^{2/3}$.  On the other hand 
the primordial velocity dispersion decreases on average as $\delta v \sim
t^{-2/3}$, consistently  with Liouville's theorem.  When a large 
overdensity enters the non-linear regime, the particles in the vicinity of 
the overdensity fall back onto it.  This implies that the phase-space sheet 
``winds up'' in clockwise fashion wherever an overdensity grows in the 
non-linear regime.  One such overdensity is shown in Fig. 1.  Before density 
perturbations enter the non-linear regime, there is only one value of 
velocity, i.e. one single flow, at a typical location in physical space,
because the phase-space sheet covers physical space only once.  On the other 
hand, inside an overdensity in the non-linear regime, the phase-space sheet 
covers physical space multiple times implying that there are several, but 
always an odd number of, flows at such locations in physical space.  

At the boundary between two regions one of which has $n$ flows and the 
other $n + 2$ flows, the physical space density is very large because the
phase-space sheet has a fold there.  At the fold, the phase-space sheet is 
tangent to velocity space and hence, in the limit of zero velocity
dispersion $(\delta v = 0)$, the density diverges since it is 
the integral of the phase-space density over velocity space.  The 
structure associated with such a phase-space fold is called a 'caustic'.
It is easy to show (see section II) that,  in the limit of zero velocity 
dispersion, the density diverges as $d \sim {1\over\sqrt{\sigma}}$ when 
the caustic is approached from the side with $n+2$ flows, where $\sigma$ 
is the distance to the caustic.  If the velocity dispersion is small but 
non-zero, the divergence is cut off so that the density is no longer 
infinite at the caustic but merely very large.  
 
In discussing this type of phenomenon, it is useful to adopt a 
parametrization of the phase-space sheet in the limit of zero 
velocity dispersion, by giving each particle a 3-parameter label 
$\vec{\alpha} = (\alpha_1, \alpha_2, \alpha_3)$.  The phase-space sheet 
location at time $t$ is specified by the map $\vec{\alpha} 
\rightarrow \vec{x} (\vec{\alpha},t)$ where $\vec{x}$ is the position in 
physical space of particle $\vec{\alpha}$ at time $t$.  The velocity of 
particle $\vec{\alpha}$ is $\vec{v} = {\partial \vec{x}\over 
\partial t}~(\vec{\alpha}, t)$.  The density diverges wherever the map 
$\vec{\alpha} \rightarrow \vec{x}$ is singular, i.e. where the Jacobian 
$D \equiv~det~\left({\partial \vec{x}\over \partial \vec{\alpha}}\right)$ 
vanishes.  Thus caustics are associated with zeros of $D$.  Since $D=0$ is 
one condition on three parameters, caustics are generically two dimensional 
surfaces.

Y. Zel'dovich \cite{Zel'} emphasized the importance of caustics in large 
scale structure formation, suggesting the name ``pancakes'' for them.  The 
reason why galaxies tend to lie on surfaces \cite{Huchra}, such as ``the 
Great Wall'', is undoubtedly that the 3D sheet on which the dark matter 
particles and baryons lie in phase-space acquires folds on very large 
scales, producing caustics appropriately called ``Zel'dovich pancakes''.  

S. Tremaine \cite{Tremaine} recently used the techniques of Catastrophe 
Theory \cite{Gilmore} to catalogue the caustics which may occur in 
observations of structure formation.

We saw above that where a localized overdensity is growing in the 
non-linear regime, the line which is at the intersection of the phase-space 
sheet with the $(z,\dot{z})$ plane winds up in clockwise fashion.  The onset 
of this process is illustrated in Fig. 1.  Of course, the picture is 
qualitatively the same in the $(x,\dot{x})$ and $(y,\dot{y})$ planes.  In 
this view, the process of galactic halo formation is the winding up of the 
phase-space sheet of collisionless dark matter particles.  When the 
galactic center is approached from any direction, the local number of 
flows increases.  First, there is one flow, then three flows, then five, 
seven...  The number of flows at our location in the Milky Way galaxy has 
been estimated to be of order 100 \cite{is}.  The boundary between the
region with one (three, five, ...) and the region with three (five, seven,
...) flows is the location of a caustic which is topologically a sphere 
surrounding the galaxy.  When these caustic spheres are approached from 
the inside the density diverges as $d\sim {1\over\sqrt{\sigma}}$ in the 
zero velocity dispersion limit.  I call these spheres ``outer'' caustics to 
distinguish them from the ``inner'' caustics which are the main topic of this 
paper.

To see inner caustics, let us first discuss the case where the overdensity 
is spherically symmetric and all the dark matter particles carry zero 
angular momentum with respect to the center.  All particles move on radial 
orbits then.  As a result the galactic center is a caustic point where the 
density associated with each flow in and out of the galaxy diverges as 
$d\sim {1\over r^2}$ in the limit of zero velocity dispersion, where $r$ is 
the radial coordinate.  We remarked earlier that a caustic is generically 
a surface.  Thus we find a caustic {\it point} only because we are assuming 
spherical symmetry and purely radial orbits.  If these assumptions are relaxed, 
the caustic at the galactic center will spread over some size $a$.  The 
notion of caustic point is nonetheless useful provided $a$ is small enough.  
Indeed, for $r>> a$ the density will still behave as $d\sim {1\over r^2}$ and 
the phase-space structure will be qualitatively the same as for $a=0$. 

The question arises what is the inner caustic in the absence of spherical 
symmetry and in the presence of angular momentum.  I. Tkachev, Y. Wang and 
I found \cite{sty,me} that the inner caustic is a ring, i.e. a closed line.  
The density diverges as $d\sim {1\over \sigma}$ in the zero velocity 
dispersion limit where $\sigma$ is the distance to the line \cite{me}.  
However, as remarked earlier, a caustic is generically a surface.  So a 
caustic line must again be a special degenerate case.  Indeed, we will see 
below that a caustic ring is more precisely a closed tube with a special 
structure.  Nonetheless, when the transverse dimensions, called 
$p$ and $q$ below, are small compared to the radius of curvature $a$ of 
the tube, $d\sim {1\over\sigma}$ for $a>> \sigma>> p,q$.  Hence, it makes 
sense to think of the caustic ring first as a closed line and then, on 
closer inspection, as a closed tube.  

The next question is what is the structure of the tube.  We find below 
that its transverse cross-section is a closed line with three cusps, one 
of which points away from the galactic center (see Figs. 5 and 6).  In the 
language of Catastrophe Theory such a singularity is called a $D_{-4}$ 
catastrophe \cite{Gilmore}.  It has a triality (i.e. a $Z_3$ invariance) 
which is reminiscent of the triality of the Lie group $D_4$, also called 
SO(8).

The existence of caustic rings of dark matter results from only two 
assumptions (see section III for a derivation):
\begin{enumerate}
\item the existence of collisionless dark matter
\item that the velocity dispersion of the infalling dark matter is much 
less, by a factor ten say, than the rotation velocity of the galaxy.
\end{enumerate}

\noindent Only the second assumption requires elaboration.  We noted 
earlier that velocity dispersion smoothes out caustics.  The question is 
when is the velocity dispersion so large as to smooth caustic rings over 
distance scales of order the ring radius $a$, thus making the notion 
of caustic ring meaningless.  In ref. \cite{me} this critical velocity 
dispersion was estimated to be 30 km/s = $10^{-4}$ for the caustic rings 
in our own galaxy, whose rotation velocity is 220 km/s.  $10^{-4}$ is much 
less than the {\it primordial} velocity dispersion $\delta v$ of the cold 
dark matter candidates.  However the velocity dispersion $\Delta v$ 
associated with density perturbations also smoothes caustics in coarse 
grained observations.  From the point of view of an observer with infinite 
resolution, the effect of $\Delta v$ on a caustic surface is to make it 
bumpy.  However, to an observer with poor spatial resolution, this is the 
same as smoothing the surface.  So the question is whether the velocity 
dispersion $\Delta v$ of cold dark matter particles associated with 
density perturbations falling onto our galaxy is less than 30 km/s.  The 
answer is yes with very high probability since the infalling dark matter 
particles are not associated with any observed inhomogeneities, and the 
velocity dispersion of an inhomogeneity as large as the Large Magellanic 
Cloud is still only 10 km/s.  The only way the velocity dispersion of the 
infalling dark matter particles could be as large as say 20 km/s is for 
these particles to be part of clumps whose mass/size ratio is 4 times 
larger than that of the Large Magellanic Cloud.  But if that were the case, 
why did these clumps fail to become luminous? 

I have interpreted \cite{me} the appearance of bumps in the rotation 
curves of NGC3198 and of our own galaxy as due to caustic rings of 
dark matter.  That interpretation makes use of an additional assumption, 
namely that the infall of dark matter particles is self-similar
\cite{ss,sty2,sty}.  This additional assumption is not used in this 
paper.  The goal is to clearly distinguish those conclusions which 
follow exclusively from the two assumptions listed above from those 
conclusions which require the extra assumption of self-similarity.  We 
are however motivated by the possiblility that caustic rings may be 
observed as bumps in rotation curves and we derive the shape that such 
a bump should ideally have.  Gravitational lensing by caustic rings
\cite{Hogan} is another technique by which these structures may be 
observed.

One might ask whether caustic rings can be seen in N-body simulations 
of galaxy formation.  Zel'dovich pancakes are seen \cite{Melott}. 
However, caustic rings would require far greater resolution than 
presently available, at least in a 3D simulation of our own halo.  
Indeed, the largest ring in our galaxy has been estimated to have 
radius of order 40 kpc \cite{me}.  It occurs in a flow that extends 
to the Galaxy's current turnaround radius, of order 2 Mpc.  To resolve 
this first ring, the spatial resolution would have to be of order 
10 kpc or smaller.  Hence a minimum of 
$2\cdot{1 \over (10{\rm kpc})^3} {4\pi \over 3}(2 {\rm Mpc})^3
\simeq 7\cdot 10^7$ particles would be required to see the caustic 
ring in a simulation of this one flow.  However, the number of flows 
at 40 kpc in our halo is of order 10 \cite{sty}.  So it appears that 
several times $10^8$ particles are necessary in a 3D simulation of our 
halo.  This is a strict minimum because it only addresses the kinematic 
requirement of resolving the halo in phase-space, assuming moreover 
that the particles are approximately uniformly distributed on the 
phase-space sheets.  There is a further dynamical requirement that 
2-body collisions do not artificially 'fuzz up' the phase-space sheets.  
Indeed 2-body collisions are entirely negligible in the flow of cold 
dark matter particles such as axions or WIMPs.  As a result, 
Liouville's theorem is strictly obeyed.  On the other hand, 2-body 
collisions are present in N-body simulations and thus Liouville's 
theorem is violated.  As a result the velocity dispersion is 
artificially increased in the simulations.  This may occur to such 
an extent that the caustics are washed away even if several $10^8$ 
particles are used.  The best bet to see a caustic ring would obviously 
be in a 2D simulation of an axially symmetric flow.   

This paper is organized as follows.  In section II, we give a general 
discussion of caustic surfaces and caustic lines.  Caustic lines are 
a degenerate case whereas caustic surfaces are generic.  We distinguish  
two types of caustic line which we call {\it attached} and {\it isolated}.
The former are attached to caustic surfaces whereas the latter are not.
Caustic rings are of the {\it isolated} type.  In section III, we discuss 
the caustics associated with the infall of collisionless dark matter 
particles onto a galaxy.  We focus most of our attention on the inner 
caustic rings.  We show that the presence of these rings follows only 
from the two assumptions listed above.  We find that the ring is a tube.  
Inside the tube are four flows whereas outside the tube are two flows.  
(In addition there is an odd number of flows which are not associated 
with the caustic ring.)  In section IV, we give a detailed analysis 
of the caustic ring under the additional assumptions that the flow 
is axially and reflection symmetric and that the transverse 
dimensions of the ring, $p$ and $q$, are small compared to the 
ring radius $a$.  In section IVA, we show that under these assumptions
the flow near the ring is described in terms of five parameters: $a$, 
$b$, $\tau_0$, $u$ and $s$.  In section IVB, we relate these parameters 
to the velocity distribution of the infalling dark matter particles.  In 
section IVC, we give a qualitative description of the flow of particles 
on distance scales of order the ring radius $a$.  In section V, the 
results of section IV are used to derive the shape of the bump a caustic 
ring causes in a galactic rotation curve if the ring lies in the galactic 
plane and the rotation curve is measured in the galactic plane.  Section VI
summarizes the conclusions.

\section{Caustics in general}

Consider a flow of collisionless particles with zero (or negligible) 
velocity dispersion.  Being collisionless, the particles obey 
Liouville's theorem.  Since they have negligible velocity dispersion,
the particles lie on a time-dependent 3D sheet in 6D phase-space.  The 
flow is completely specified by giving the spatial coordinates 
$\vec {x} (\vec \alpha, t)$ of the particle labeled 
$\vec \alpha$ at time $t$, for all $\vec\alpha$ and $t$.  The 3-parameter 
label $\vec\alpha$ is chosen arbitrarily.  Let $\vec\alpha_j (\vec x, t)$, 
with $j=1...n$, be the solutions of $\vec x = \vec x (\vec\alpha, t)$.  
$n$ is the number of distinct flows at $\vec x$ and $t$.  The total number 
of particles is:
\begin{equation}
N = \int d^3\alpha ~{d^3N\over d\alpha_1 d\alpha_2 d\alpha_3} 
~(\vec\alpha) 
= \int d^3x \sum_{j=1}^n ~{d^3N\over d\alpha_1 d\alpha_2 d\alpha_3} 
~\left(\vec\alpha_j (\vec x,t)\right) {1\over 
\mid det \left({\partial \vec x\over \partial 
\vec\alpha}\right)\mid_{\vec\alpha_j (\vec x, t)}} \ .
\label{2.1}
\end{equation}
The density of particles in physical space is thus:
\begin{equation}
d(\vec x,t) = \sum_{j=1}^n~{d^3N\over d\alpha_1 d\alpha_2 d\alpha_3}~
(\vec\alpha_j (\vec x,t))~{1\over \mid D(\vec\alpha,t)
\mid_{\vec\alpha_j (\vec x, t)}}~~~~~~\ , 
\label{2.2}
\end{equation}
where
\begin{equation}
D(\vec\alpha, t) \equiv det \left( {\partial \vec x\over \partial 
\vec\alpha}\right)\ .\label{2.3}
\end{equation}
The formula for the density is, of course, reparametrization invariant.
Caustics occur wherever $D=0$, i.e. where the map $\vec\alpha 
\rightarrow \vec x$ is singular.  At the caustic the density diverges.  
The divergence is cut off if the velocity dispersion is finite.

Generically the zeros of $D$ are simple, i.e. the matrix 
${\partial \vec x\over \partial \vec\alpha}$ has a single vanishing 
eigenvalue.  The condition that one eigenvalue vanishes imposes one 
constraint on the three parameters $\vec{\alpha}$.  Hence a caustic 
is generically a 2D surface in physical space.

\subsection{Generic surface caustics}

Consider a generic surface caustic at a fixed time $t$.  We may 
reparametrize the flow near the caustic, 
$\vec\alpha \rightarrow \vec\beta = \vec\beta (\vec\alpha, t)$, such 
that the caustic surface is at $\beta_3 =0$.  Also, in a neighborhood 
of a point on the surface, choose Cartesian coordinates such that 
$\hat{z}$ is perpendicular to the surface whereas $\hat{x}$ and 
$\hat{y}$ are parallel.  We have then:
\begin{equation}
D = {\partial z\over \partial\beta_3} ~det\left({\partial (x,y)\over 
\partial(\beta_1,\beta_2)}\right)
\label{2.4}
\end{equation}
near that point.  The 2-dim. matrix ${\partial (x,y)\over
\partial(\beta_1,\beta_2)}$ is non-singular.  Since $D=0$ at $\beta_3 = 0$, 
we have
\begin{equation}
z = z_0 + B\beta_3^2
\label{2.5}
\end{equation}
for small $\beta_3$.  We may orient the $\hat{z}$-axis in such a way that 
$B>0$.  Then
\begin{equation}
D = 2\sqrt{B(z-z_0)}~det \left({\partial (x,y)\over \partial (\beta_1, 
\beta_2)}\right)~~~~~{\rm for}~z> z_0 \ .
\label{2.6}
\end{equation}
Hence, near a caustic surface located at $z = z_0$, the density diverges 
as ${1\over \sqrt{z-z_0}}$ on one side of the surface.

Fig. 2a shows a 2D cut of phase-space along the $(z,\dot{z})$ plane.  The 
particles lie on a line which is at the intersection of the phase-space 
sheet with the $(z,\dot{z})$ plane.  The label $\beta_3$ gives the 
position of the particles along the line.  In the example of the figure, 
there are two caustic surfaces, one at $z = z_1$ and the other at 
$z = z_2$.  The two dimensions ($x$ and $y$) into which the caustic 
extends as a surface are not shown.  The density $d(z)$, shown 
in Fig. 2b,  diverges as ${1\over\sqrt{z-z_1}}$ for $z\rightarrow z_1$ 
with $z>z_1$ and as ${1\over\sqrt{z_2-z}}$ for $z\rightarrow z_2$ with 
$z<z_2$.  For $z_1 < z < z_2$ there are three flows ($n=3$) whereas for 
$z <z_1$ and $z > z_2$ there is only one flow ($n=1$).  The phase-space 
sheet ``folds back'' at $z=z_1$ and $z=z_2$.  That is why the map 
$\beta \rightarrow z$ is singular at these locations.  

\subsection{Line caustics}

Next, consider places where $D$ has a double zero, i.e. where 
${\partial\vec x\over \partial\vec \alpha}$ has vanishing eigenvalues
for two different eigenvectors.  The condition that two eigenvalues vanish 
defines a line.  This line is the location of a more singular kind of 
caustic which we call 'line caustic'.

In a small neighborhood of a line caustic let us reparametrize the flow 
$(\vec\alpha \rightarrow \vec\beta)$ such that $\beta_1 =\beta_2=0$ defines 
the line in the new coordinates.  Also, choose Cartesian coordinates such 
that $\hat{z}$ is parallel to the line and $\hat{x}$ and $\hat{y}$ are 
perpendicular to it.  In this neighborhood we have:
\begin{equation}
D = {\partial z\over \partial\beta_3}~det~\left({\partial (x,y)\over 
\partial (\beta_1,\beta_2)}\right)\label{2.7}
\end{equation}
with ${\partial z\over \partial\beta_3} \neq 0$.  At $\beta_1 = \beta_2 = 0$, 
the matrix ${\partial (x,y)\over \partial (\beta_1, \beta_2)}$ vanishes 
since it is 2x2 and has vanishing eigenvalues for two different eigenvectors.
Thus for small $\beta_1$ and $\beta_2$,
\begin{eqnarray}
x &=& x_0 + {1\over 2} X_{11}\beta_1^2 + X_{12}\beta_1\beta_2 + {1\over 2} 
X_{22} \beta_2^2 + 0 (\beta^3) \nonumber \\
y &=& y_0 + {1\over 2} Y_{11} \beta_1^2 + Y_{12} \beta_1 \beta_2 + {1\over 
2} Y_{22} \beta_2^2 + 0(\beta^3)\ .
\label{2.8}
\end{eqnarray}
Linear combinations $x - x_0 + \gamma (y - y_0)$ are complete squares provided:
\begin{equation}
\gamma^2 (Y_{11} Y_{22} - Y_{12}^2) + \gamma (Y_{11} X_{22} + X_{11} 
Y_{22} - 2 X_{12} Y_{12}) + X_{11} X_{22} - X_{12}^2 = 0\ ,
\label{2.9}
\end{equation}
which has two solutions:
\begin{eqnarray}
\gamma_\pm &=& {1\over 2(Y_{11} Y_{22} - Y_{12}^2)} \bigl\{ - Y_{11} 
X_{22} - X_{11} Y_{22} + 2 X_{12} Y_{12} \bigr.\nonumber \\
&\pm& \bigl.\left[\left(Y_{11}X_{22}+X_{11}Y_{22} - 2X_{12}Y_{12}\right)^2
- 4 \left(Y_{11} Y_{22} - Y_{12}^2\right) \left( X_{11} 
X_{22} - X_{12}^2\right) \right]^{1/2} \bigr\}~~\ .
\label{2.10}
\end{eqnarray}
If the denominator $Y_{11} Y_{22} - Y_{12}^2 = 0$, one should
interchange the role of $x$ and $y$.  If $Y_{11} Y_{22} - Y_{12}^2 = 0$ 
and $X_{11} X_{22} - X_{12}^2 = 0$, $x - x_0$ and $y-y_0$ are complete 
squares to begin with.  Note that Im$\gamma_\pm$ may differ from zero 
and that $\gamma_- = \gamma_+^\ast$ in this case.  We have:
\begin{equation}
x_\pm \equiv x - x_0 + \gamma_\pm (y - y_0) 
= {1\over 2} X_\pm \left( \beta_1 + \eta_\pm \beta_2\right)^2
\label{2.11}
\end{equation}
with
\begin{equation}
X_\pm = X_{11} + \gamma_\pm Y_{11}\label{2.12.a}
\end{equation}
and
\begin{equation}
\eta_\pm = {X_{12} + \gamma_\pm Y_{12}\over X_{11} + \gamma_\pm Y_{11}}~~\ .
\label{2.12.b}
\end{equation}
In terms of these new quantities:
\begin{equation}
D_2 \equiv det \left( {\partial (x,y)\over \partial (\beta_1, 
\beta_2)}\right) = 2\sqrt{X_+ X_- x_+ x_-}~{\eta_+ -\eta_-\over\gamma_+ - 
\gamma_-}~~~\ .
\label{2.14}
\end{equation}
There are two cases to consider, depending on whether Im$\gamma_\pm \neq 
0$ or Im$\gamma_\pm = 0$.

If Im$\gamma_\pm \neq 0$, let $\gamma_\pm \equiv \gamma_1 \pm i\gamma_2,
~X_\pm \equiv X_1 \pm i X_2,~x_\pm \equiv x_1 \pm i x_2$ and 
$\eta_\pm \equiv \eta_1 \pm i \eta_2$.  $\gamma_1, \gamma_2, X_1, X_2,
x_1, x_2, \eta_1$ and $\eta_2$ are real.
We have:
\begin{equation}
D_2 = 2~{\eta_2\over \gamma_2} ~ \sqrt{X_1^2 + X_2^2} \left[\left(x-x_0 + 
\gamma_1 (y-y_0)\right)^2 + \gamma_2^2 (y-y_0)^2\right]^{1/2}\ 
.\label{2.15}
\end{equation}
$D_2 \neq 0$ everywhere except at $(x,y) = (x_0, y_0)$.  The density 
$d\sim {1\over \sigma}$ with
\begin{equation}
\sigma \equiv \sqrt{\left(x - x_0 + \gamma_1 (y - y_0)\right)^2 + 
\gamma_2^2 (y - y_0)^2}\ .\label{2.16}
\end{equation}
An illustrative example of a Im$\gamma_\pm \neq 0$ line caustic is:
\begin{equation}
x = x_0 + X {1\over 2} (\beta_1^2 - \beta_2^2)~~,~~
y = y_0 + Y \beta_1\beta_2
\label{2.19}
\end{equation}
for which $D = 2 \sqrt{Y^2(x-x_0)^2 + X^2(y-y_0)^2}$.  We call a 
Im$\gamma_\pm \neq 0$ line caustic "isolated" because it is not 
attached to any surface caustic.  Ring caustics associated with the 
infall of collisionless dark matter onto a galaxy are line caustics of 
the isolated type.  They will be discussed in detail in sections III 
and IV.

If Im$\gamma_\pm = 0$, then $x_\pm,~ X_\pm$ and $\eta_\pm$ are real.  From 
Eq. (2.11) and Eq. (2.14) we see that $D$ has simple zeros along two surfaces, 
$x_+ = 0$ and $x_- = 0$, and a double zero on the line where these two 
surfaces meet.  Thus the Im$\gamma_\pm = 0$ line caustic is "attached"
to two generic surface caustics.  The following is an illustrative example:
\begin{equation}
x = x_0 + {1\over 2} X\beta_1^2, ~~~~
y =  y_0 + {1\over 2} Y \beta_2^2\ .
\label{2.17}
\end{equation}
In this case $D = 2 \sqrt{XY(x-x_0)(y-y_0)}$, and hence the density 
$d\sim {\theta (x-x_0)\theta (y-y_0)\over \sqrt{(x-x_0)(y-y_0)}}$ for 
$X,Y>0$.  We have no use for the "attached" type of line caustic in this
paper.

\section{Caustic rings}

In this section we give a general discussion of the caustic rings that are 
associated with the infall of collisionless dark matter onto a galaxy.  
No symmetry is assumed.  We do assume that the velocity dispersion of the 
dark matter particles is small compared to the rotation velocity of the 
galaxy. We set the velocity dispersion equal to zero and derive the 
existence of caustics.  A small velocity dispersion provides a cutoff, 
so that the density at the caustic does not become infinite but merely 
very large.

Fig. 3 shows successive time frames of a set of collisionless particles 
falling through a galaxy.  The particles move purely under the effect of 
gravity.  In Figure 3a, the particles are at first 'turnaround',  i.e. they 
are about to fall onto the galaxy for the first time in their history.  They 
are located on a closed 2D surface surrounding the galaxy in physical space, 
called the ``turnaround sphere''.  Fig. 3 shows the intersection of this 
sphere with the plane of the figure, as it evolves in time.  For the sake 
of definiteness it is assumed that the particles carry net angular momentum 
about the vertical axis.  Qualitatively speaking, the turnaround sphere is 
spinning about the vertical.  The particles near the top (bottom) of the 
sphere in frame a carry little angular momentum and end up near the bottom 
(top) of the sphere in frame f after falling through the galaxy.  The 
particles near the equator carry the most angular momentum.  They form a 
ring whose radius decreases in time down to some minimum value, reached near 
frame d, and then increases again.  Generally the radius of the sphere at 
second turnaround is smaller than at first turnaround because the galaxy 
has grown by infall in the meantime.  After second turnaround the sphere 
falls back in and repeats the same qualitative sequence till third turnaround, 
and so on.

There is a generic surface caustic associated with the $m$th turnaround 
where $m=2,3,4, ...$.  These caustics are located near where the particles 
of a given outflow reach their maximum radius before falling back in.   
To see this, parametrize the flow at a given time $t$ by $\vec{x} 
(\alpha,\beta, t_0;t)$ where $t_0$ is the time the particle was at last 
turnaround and $\alpha,\beta$ -- for example, spherical coordinates -- 
tell us where the particle was on the turnaround sphere at that time.  We
define:  $\vec{x}^0 \equiv {\partial \vec{x}\over \partial 
t_0}~,~\vec{x}^1 \equiv {\partial \vec{x}\over \partial\alpha}$, 
$\vec{x}^2 \equiv {\partial\vec{x}\over\partial\beta}~,$ and 
$\dot{\vec{x}} \equiv {\partial\vec{x}\over\partial t}$.  In terms of 
these:
\begin{equation}
D =~\vec{x}^0~\cdot~(\vec{x}^1\times \vec{x}^2)\ .
\label{3.2n}
\end{equation}
Since the discussion is for a fixed time $t$, let us not show the $t$
dependence explicitly further.  Let us assume that 
$\vec{x} (\alpha,\beta, t_0)$ is near particles at 
their $m$th turnaround with $m = 2,3,4...$  The spheres $\{\vec{x} 
(\alpha,\beta, t_0^\prime)~:~\forall\alpha,\beta\}$ with $t_0^\prime$ 
considerably larger or considerably smaller than $t_0$ are inside the 
sphere $\{\vec{x} (\alpha,\beta, t_0)~:~\forall\alpha,\beta\}$.  This 
implies that for all $\alpha,\beta$ there is a $t_0 (\alpha,\beta)$ such 
that $\vec{x}^0 (\alpha,\beta, t_0 (\alpha,\beta))$ is parallel to the 
sphere $\{\vec{x} (\alpha^\prime, \beta^\prime, t_0 (\alpha,\beta)~:~ 
\forall\alpha^\prime,\beta^\prime\}$, and hence where $D=0$.  Therefore 
the sphere 
$\{\vec{x} (\alpha,\beta,t_0 (\alpha,\beta))~:~\forall\alpha,\beta\}$ 
is the location of a generic surface caustic.  These {\it outer} 
caustics were mentioned in the introduction as the boundaries between 
the regions with $n$ flows and those with $n+2$ flows where $n=1,3,5 ...$
We do not discuss them further in this paper.  Our focus is upon the 
{\it inner} caustics which, as we will soon see, have the shape of rings 
(closed tubes) and which are located near where the particles with the most 
angular momentum in a given inflow reach their distance of closest approach 
to the galactic center before moving back out of the galaxy.

To see inner caustic rings, let us return to the sphere of Fig. 3.  During 
each infall-outfall sequence, the sphere turns itself inside out.  Indeed, 
a particle which is part of the flow and is just inside the sphere in frame 
b of Fig. 3 is outside the sphere in frame e, and vice-versa.  There is 
therefore a ring of points in space-time which are inside the sphere last. The 
intersection of this space-time ring with the plane of the figure is at two 
space-time points, one located at the cusp in frame d, the other at the cusp 
in frame e.  Since there is a continuous flow of spheres falling in and out, 
the ring just defined is a persistent feature in space.   We now show that it 
is the location of a caustic.

Fig. 4 shows the infall sphere in a neighborhood of space and time where it 
is completing the process of turning itself inside out.  We choose Cartesian 
coordinates such that $\hat{y}$ is parallel to the ring at that point; 
$\hat{x}$ and $\hat{z}$ are as shown.  We parametrize the flow in a small 
neighborhood of the ring by $\vec{\alpha} = (\alpha, \beta, t_0)$ such that 
${\partial x\over \partial\beta} = {\partial z\over \partial\beta} = 0$.  
As before, $t_0$ labels successive infall spheres and may be taken to  be
the time of their last turnaround. Thus:
\begin{equation}
D = {\partial y\over \partial\beta} \left({\partial 
x\over\partial\alpha}~{\partial z\over \partial t_0} - {\partial x\over 
\partial t_0}~{\partial z\over \partial\alpha}\right)\ .
\label{3.1}
\end{equation}
Fig. 4a shows an infall sphere just before it reaches the ring.  The line 
is the intersection of the infall sphere with the plane of the figure.  
$\alpha$ labels points along the line.  We have 
${\partial z\over\partial\alpha} = 0$ at points A and B because the line 
is parallel to the $\hat{x}$-axis at these points.  Similarly, 
${\partial x\over\partial\alpha} = 0$ at point C.  Fig. 4b shows an infall 
sphere at the moment it reaches the ring.  Points A, B and C in Fig. 4a 
have moved to point E in Fig. 4b.  Hence 
${\partial z\over \partial\alpha} = {\partial x\over\partial\alpha} = 0$ 
and therefore $D=0$ at point E, which is thus the location of a  caustic.

In general, $D$ only has a simple zero at E, which means that E is the 
location of a {\it surface} caustic.  We show below that the complete 
inner caustic surface is a closed tube.  E is a point on this tube.  In 
the limit where the transverse size $p$ of the tube goes to zero, both
${\partial z\over \partial\alpha} = {\partial x\over\partial\alpha} = 0$
and ${\partial x\over \partial t_0} = {\partial z\over\partial t_0} = 0$
at E.  $D$ then has a double zero at E.  In that limit the tube caustic
becomes an {\it isolated} line caustic where the density diverges as 
${1\over\sigma}$.  When $p$ is finite but small, the density 
$d(\sigma)\sim {1\over \sigma}$ for $\sigma >> p$ but has a more 
complicated behaviour for $\sigma\sim p$.

Let us show that the inner caustic must have the topology of a tube, 
starting with the case of Fig. 3 and generalizing from there.  In Fig. 3b, 
$\vec{x}^0$ (not shown explicitly) is everywhere pointing outward of the 
infall sphere because later infall spheres are outside this one.  In Fig. 3f, 
$\vec{x}^0$ is pointing inside because later infall spheres are inside this 
one.  This implies that during the infall $\vec {x}^0$ either vanishes at 
some space-time points or becomes parallel to the sphere.  Since $D=0$ at 
such points, they are the location of caustics.  Since $D$ is a continuous 
function of $\vec{\alpha}$, the caustic must lie on a closed surface.
Now, if we follow the motion of points which are near the top (bottom) of 
the sphere in Fig. 3b and end up near the bottom (top) in Fig. 3e, 
$\vec{x}^0$ always points up (down).  $\vec{x}^0$ does not vanish and is 
not parallel to the sphere at any time between these two frames for these 
points.  This implies that there are no inner caustics within some cylinder 
extending from top to bottom in the spatial volume under consideration in 
Fig. 3.  On the other hand, for the points near the equator in Fig. 3, 
$\vec{x}^0$ is pointing outward during infall and is pointing inward during 
outfall.  Thus if we track a point near the equator, at some time $\vec{x}^0$ 
either vanishes or is parallel to the sphere.  The points where 
this happens lie on a closed surface which is outside the previously 
defined cylinder but wraps around it.  That surface must therefore be a tube.
The tube is located near the equator, where the particles with the most 
angular momentum are at their distance of closest approach.

Consider now the most general angular momentum distribution on the 
turnaround sphere.  The angular momentum is a continuous two-dimensional 
vector field on the sphere.  It is well-known that such a vector field 
must have (at least) two zeros.  If we track the particles in the neighborhood  
of these angular momentum zeros, we find that their $\vec{x}^0$ does not 
vanish and is not parallel to the sphere at any time during the infall-outfall
sequence.  Therefore the inner caustic appears only in the flow of particles 
which are some distance away from both zeros.  Since that set of particles
has the topology of a closed ribbon, and the previously defined caustic 
ring (the set of points E which are in the turnaround sphere last) goes
around this closed ribbon once, the inner caustic must be a closed surface 
with one handle, i.e. a closed tube.  Note that the closed tube may 
self-intersect and may have a complicated shape.  However, if the angular 
momentum is dominated by a smooth component which carries net angular 
momentum, as was assumed in Fig. 3, then the tube resembles a circle.  
If there is no angular momentum at all, the tube reduces to a point at the 
galactic center.  As was mentioned in the Introduction, the galactic center 
is a point caustic in that case, where the density behaves as 
$d(r)\sim {1 \over r^2}$.

Fig. 5 shows simultaneous (same $t$) infall spheres near the caustic 
corresponding to five different initial times:  
$t_{01} > t_{02} > ... > t_{05}$.  The five numbered dots show the positions 
$\vec{x}(\alpha,t_{0k}), k = 1...5$, for some fixed $\alpha$ and the five 
different initial times.  Following the dots gives one a qualitative 
picture of the flow in time.  The $t_{01}$ sphere is falling in but has not 
yet crossed itself, as in frame b of Fig. 3.  The $t_{02}$ sphere has crossed 
itself but has not yet completed the process of turning itself inside out, 
as in frame c of Fig. 3.  The $t_{03}$ sphere is just completing the process 
of turning itself inside out, as in frame d of Fig. 3.  The cusp at point E 
is the location of a caustic for the reason given earlier.  Let $\alpha_E$
be the value of $\alpha$ at E.  Thus $\vec{x}(\alpha_E,t_{03})$ is the 
position of E.  In the example of the figure, the particles at the cusp are 
moving to the left.  Thus at $(\alpha_E, t_{03})$, $\dot{\vec{x}}$ is 
pointing to the left whereas $\vec{x}^0$ is pointing to the right of the 
$t_{03}$ infall sphere.  For smaller initial times, such as $t_{05}$, 
$\vec{x}^0$ is pointing to the left of the infall sphere.  Let $t_{04}$ be 
the initial time and F be the point where $\vec{x}^0(\alpha_E,t_0)$ crosses 
the $t_0$ sphere.  In view of Eq.~(3.1), F is the location of a caustic
as well.  Consider any point which is far from both E and F.  At such a point 
there are two flows because the sphere passes such a point twice, once on 
the way in and once on the way out.  Consider also a point located between 
E and F.  At such a point there are four flows because the sphere passes by 
four times:  at fixed $t$, twice for initial time $t_0$ between $t_{02}$ and 
$t_{03}$, once between $t_{03}$ and $t_{04}$ and once between $t_{04}$ and 
$t_{05}$.  There is therefore a finite compact region in the plane of Fig. 5 
inside of which there are four flows and outside of which there are two flows.  
The boundary of this region, shown as a closed dashed line with three cusps, 
is the location of the tube caustic.  The reason for the three cusps will be 
given in the next section.  

In the limit where the tube shrinks to a line, $D$ has a double zero on the
line.  Indeed E and F coincide in that limit.  $\vec{x}^0$ is parallel to   
the ring at E because $\vec{x}^0$ is parallel to the infall sphere at F which 
coincides with E and the infall sphere has a cusp at E.  Eq.~(3.2) shows that 
$D$ has a double zero then which means that the ring is a line caustic.  It 
is of the isolated type since it is not attached to any surface caustics.

In the example of the figure, the initial time $t_{03}$ of E is after the 
initial time $t_{04}$ of F.  The opposite is equally possible.  Indeed the 
time reverse of the sequence of Fig. 5 is also a possible sequence and it 
has the times of E and of F in the opposite order.  On the other hand, the
density is invariant under time reversal of the flow.  This shows that point 
F is always closer to the galactic center than point E.  Note also that if 
the flow is time reversal invariant (i.e. if the outfall sequence is the time 
reverse of the infall sequence), the tube caustic collapses to a line caustic 
because $t_{04} = t_{03}$ and hence E and F coincide.

If the successive spheres all fall in exactly the same way, the caustic 
ring is stationary.  In general, however, the trajectories of successive 
spheres change with time $t$, albeit slowly.  As a result the caustic 
ring moves about.  However, whether or not it moves, the caustic ring 
is perfectly sharp in the limit of zero velocity dispersion.

\section{The axially symmetric case}

We assume in this section that the flow of dark matter in and out of 
the galaxy is axially symmetric about $\hat{z}$ and symmetric under  
reflection $z\rightarrow -z$.  We also assume that the transverse 
dimensions, called $p$ and $q$ below, of the ring caustic are small 
compared to the ring radius $a$.  As before, we assume that the dark
matter particles are collisionless and we neglect their velocity
dispersion.  We use the following parametrization of the flow.  
Let $R(t_0)$ be the turnaround radius in the $z=0$ plane at time $t_0$.  
Then let $\vec{x} (\theta_0, \varphi_0, t_0; t)$ be the position at time 
$t$ of the particle that was  at the location of polar coordinates 
$(\theta_0, \varphi_0)$ on the sphere of radius $R(t_0)$ at time $t_0$.  
The density is given by Eqs. (2.2-3) with 
$\vec{\alpha} = (\theta_0, \varphi_0, t_0)$ and $n = 2(4)$ outside (inside) 
the closed tube caustic described in a general fashion in the previous 
section.  

However, since we assume axial symmetry the motion is in effect 
two-dimensional.  Let $\rho(\alpha, t_0; t)$ and $z(\alpha, t_0; t)$ be 
the cylindrical coordinates at time $t$ of the ring of particles which 
start with azimutal angle $\theta_0 = {\pi\over 2} - \alpha$ at initial 
time $t_0$.  One readily shows, repeating for the axially symmetric case 
the steps which led to Eqs. (2.2-3), that the density is given by:
\begin{equation}
d(\rho,z,t) = {1\over 2\pi\rho}~\sum_{j=1}^n~{d^2N\over d\alpha 
dt_0}~(\alpha, t_0)~{1\over\mid D_2(\alpha, t_0)\mid}\Biggl|_{(\alpha, 
t_0) = (\alpha, t_0)_j} 
\label{4.1}
\end{equation}
where
\begin{equation}
D_2 (\alpha, t_0) \equiv det\left({\partial(\rho,z)\over \partial (\alpha, 
t_0)}\right)\label{4.2}
\end{equation}
and $(\alpha, t_0)_j$, with $j = 1...n$, are the solutions of 
$\rho = \rho (\alpha, t_0; t)~,~z= z(\alpha, t_0; t)$.  Under a 
reparametrization of the flow 
$(\alpha, t_0) \rightarrow [\alpha^\prime (\alpha, t_0),~t_0^\prime 
(\alpha, t_0)]$, the determinant transforms according to:
\begin{equation}
D_2 (\alpha, t_0) = D_2^\prime (\alpha^\prime, t_0^\prime)~
det\left({\partial (\alpha^\prime, t_0^\prime)\over 
\partial (\alpha, t_0)}\right)\ . 
\label{4.3}
\end{equation}
In particular,
\begin{equation}
D_2 (\alpha, t_0) = D_2^\prime (\alpha^\prime, t_0^\prime)
\label{4.4}
\end{equation}
for an $\alpha$-dependent time shift: $\alpha^\prime = 
\alpha,~t_0^\prime = t_0 + \Delta t_0 (\alpha)$.

\subsection{Flow at the caustic}

General characteristics of the flow near the caustic are described in 
section III.  Figure 5 constitutes a summary.  In the reflection symmetric 
case, points E and F have $z=0$ and $\alpha = 0$.  Let us reparametrize 
the flow $\alpha\rightarrow \alpha, ~t_0 \rightarrow \tau = 
t_0 + \Delta t_0 (\alpha)$ such that $z (\alpha, \tau = 0) = 0$ for all 
$\alpha$.  We expand $\rho$ and $z$ in powers of $\alpha$ and $\tau$ 
keeping terms up to second order only. Since we assume the reflection 
symmetry:  
$z\rightarrow - z,~\rho\rightarrow \rho,~\alpha\rightarrow -\alpha$, 
and $\tau\rightarrow\tau$, we have:
\begin{equation}
z = +b\alpha\tau
\label{4.5.a}
\end{equation}
\begin{equation}
\rho = \rho_0 - c\tau + {1\over 2} u\tau^2 - {1\over 2} s\alpha^2\ ,
\label{4.5.b}
\end{equation}
where $b, \rho_0, c, u$ and $s$ are constants.  $\rho_0$ is the 
$\rho$-coordinate of point E.  We may rewrite Eq.~(\ref{4.5.b}) as
\begin{equation}
\rho = a + {1\over 2} u (\tau -\tau_0)^2 - {1\over 2} s\alpha^2\ 
,\label{4.6}
\end{equation}
where $\tau_0 = c/u$ and $a = \rho_0 - {1\over 2}~{c^2\over u}$.  
$a$ is the $\rho$-coordinate of point F.  The parameters $b,~u$ and 
$s$ are positive.  $b$ is positive because the flow is from top to 
bottom (bottom to top) for particles with $\alpha > 0~(\alpha < 0)$.  
$u$ is positive because the particles with $\alpha = 0$ are accelerated 
outward by the angular momentum barrier.  $s$ is positive because at 
$\tau = 0$, the particles with $\alpha \neq 0$ are at 
$\rho < \rho_0$.  $\tau_0$ can either be positive or negative.  $a$ is 
always smaller than $\rho_0$, i.e. point F is always closer to the 
galactic center than point E, as noted earlier.

The determinant is
\begin{equation}
D_2 (\alpha,\tau) = - b[u\tau (\tau -\tau_0) + s\alpha^2]\ .
\label{4.7}
\end{equation}
It vanishes for
\begin{equation}
\alpha = \pm \sqrt{{u\over s} \tau (\tau_0 - \tau)} \label{4.8}
\end{equation}
with $0 < \tau < \tau_0$ if $\tau_0 > 0$, and $-\tau_0 < \tau < 0$ if 
$\tau_0 < 0$.  Substituting Eq. (\ref{4.8}) into Eqs. (4.5) and (4.7), 
we find a parametric ($\tau$ = parameter) representation of the surface 
of the tube caustic:
\begin{eqnarray}
\rho &=& a + {1\over 2} u (\tau - \tau_0) (2\tau - \tau_0)\nonumber\\
z &=& \pm b\sqrt{{u\over s} \tau^3(\tau_0 - \tau)}~~~~~ \ .
\label{4.9}
\end{eqnarray}
Fig. 6 shows a cross-section of the tube.  Its  dimensions in the 
$\hat{\rho}$ and $\hat{z}$ directions are $p = {1\over 2} u\tau_0^2~~$ 
and $~~q = {\sqrt{27}\over 4}~{b\over\sqrt{us}}~p~~$ respectively.  The 
caustic has three cusps.  It is shown below that the appearance of the 
three cusps is not special to the assumed axial and reflection symmetries 
nor to the fact that we expanded only to second order in 
Eqs.~(\ref{4.5.a}) and (\ref{4.5.b}).  

First let us note that the caustic has a $Z_3$ symmetry after an 
appropriate rescaling.  Define the rescaled parameters:
\begin{equation}
T = {\tau\over\tau_0}\ ,~~~A = \sqrt{{s\over u}}~{1\over \tau_0}~\alpha\ ,
\label{4.10}
\end{equation}
and the rescaled and shifted (in the $\hat{\rho}$ direction) coordinates:
\begin{equation}
Z = {2 \over b}~\sqrt{{s \over u}}~{1\over \tau_0^2}~z~,~~~
X = {2\over u\tau_0^2}~(\rho -a) - {1\over 4}\ .
\label{4.11}
\end{equation}
In terms of these, Eqs. (\ref{4.5.a}) and (\ref{4.6}) become:
\begin{equation}
Z = 2AT~,~~~X = (T -1)^2 - A^2 - {1\over 4}\ .
\label{4.12}
\end{equation}
These relations are invariant under the discrete transformation:
\begin{eqnarray}
Z^\prime &=& - {1\over 2} Z + {\sqrt{3}\over 2} X~~,~~~
X^\prime = - {\sqrt{3}\over 2} Z - {1\over 2} X\nonumber\\
T^\prime &=& - {1\over 2} T + {\sqrt{3}\over 2} A +3/4~~,~~~
A^\prime = - {\sqrt{3}\over 2} T - {1\over 2} A + \sqrt{3}/4\ ,
\end{eqnarray}
whose cube is the identity.  In the X-Z plane, the transformation is a 
rotation by $120^o$.  It transforms the 3 cusps of the caustic into one 
another.  Hereafter, let us call the shape of Fig.~6 a ``tricusp''.
In the language of Catastrophe Theory \cite{Gilmore}, the tricusp 
is a $D_{-4}$ catastrophe. 

The apparent reason for the two cusps which are not on the
$\hat{\rho}$-axis in Fig.~6 is that ${d\rho\over d\tau}$ and 
${dz \over d\tau}$ both vanish for $\tau = {3 \over 4} \tau_0$. 
It might be thought that the simultaneous vanishing of 
${d\rho \over d\tau}$ and ${dz \over d\tau}$ is an accident
peculiar to our assumptions of symmetry and/or our limiting the 
expansion of $\rho$ and $z$ to terms of second order in $\alpha$ and 
$\tau$.  But this is not the case.  Indeed, consider the 
equation for the location of a generic caustic in 2 dimensions:
\begin{equation}
D_2(\alpha,\tau) = {\partial \rho \over \partial \alpha}
{\partial z \over \partial \tau} - {\partial \rho \over \partial \tau}
{\partial z \over \partial \alpha} = 0~~~\ .
\label{4.21n}
\end{equation}
It defines $\alpha(\tau)$ such that $[\rho(\tau) = \rho(\alpha(\tau),\tau),
z(\tau) = z(\alpha(\tau),\tau)]$ is a parametric representation of the 
caustic location.  Wherever $\rho(\tau)$ has an extremum, $z(\tau)$ may 
be expected to have an extremum as well since
\begin{equation}
{d \rho \over d \tau} = {\partial \rho \over \partial \alpha}
{d \alpha \over d \tau} + {\partial \rho \over \partial \tau} = 0 
\label{4.22n}
\end{equation}
and Eq.~(\ref{4.21n}) imply:
\begin{equation}
{dz \over d\tau} = {\partial z \over \partial \alpha}
{d \alpha \over d \tau} + {\partial z \over \partial \tau} = 0
\label{4.23n}
\end{equation}
if ${d \alpha \over d \tau}$ is finite.  The cusp at point E may appear 
to have a different origin.  Its apparent reason is that 
$z \sim \pm \tau^{3 \over 2}$ near $\tau = 0$ which is at the boundary of 
the range of $\tau$.  However, this circumstance is an artifact of the 
parametrization used.  Indeed, we saw that the three cusps are transformed 
into one another by a $Z_3$ symmetry.  As a corollary, the cusp on the 
$\hat{\rho}$-axis can be given the same parametrization as the other two.  
In conclusion, the appearance of three cusps in the cross-section of a 
ring caustic is not an accidental consequence of our simplifying 
assumptions.  Without those assumptions, the cross-section does not have 
the exact shape of Fig. 6 but it still has three cusps, at least for 
small deviations from the description given above. 

Inside (outside) the tricusp, there are four (two) flows.  For example, 
if we restrict ourselves to the $z=0$ plane, then $\alpha = 0$ or 
$\tau = 0$.  If $\alpha = 0$, then $\rho = a + {1\over 2} u (\tau - 
\tau_0)^2 > a$.  If $\tau = 0$, then $\rho = a + {1\over 2} u \tau_0^2 - 
{1\over 2} s\alpha^2 < \rho_0$.  Thus for $z=0$ and $\rho > a$, there are 
2 flows:
\begin{equation}
\alpha = 0~~~,~~~\tau = \tau_0 \pm \sqrt{{2\over u} (\rho - a)}\ ,
\label{4.14a}
\end{equation}
for which
\begin{equation}
{\partial z\over \partial \tau} = 0~~~,
~~~{\partial \rho\over \partial \tau} = \pm \sqrt{2u(\rho - a)}\ .
\label{4.14b}
\end{equation}
We may call these the ``in and out'' flows.  For $z=0$ and $\rho <\rho_0$, 
there are 2 other flows:
\begin{equation}
\tau = 0~~~,~~~\alpha = \pm \sqrt{{2\over s} (\rho_0 - \rho)}\ ,
\label{4.15a}
\end{equation}
for which
\begin{equation}
{\partial z\over \partial \tau} = \pm b \sqrt{{2\over s} (\rho_0 
-\rho)}~~~,
~~~{\partial \rho \over \partial \tau} = -u\tau_0\ .
\label{4.15b}
\end{equation}
We may call these the ``up and down'' flows.  Therefore, in the $z=0$
plane, there are four flows (in, out, up and down) for 
$a < \rho < \rho_0$, whereas there are two flows (up and down) for 
$\rho < a$, and two flows (in and out) for $\rho > \rho_0$.  

Away from the cusps, the caustic is a generic surface caustic, as
described in section IIA.  Thus, if one approaches the boundary of the 
tricusp from the inside and away from any of the cusps, the density 
increases as $d \sim {1\over\sqrt{\sigma}}$ where $\sigma$ is the distance 
to the boundary.  If the boundary is approached from the outside, away from 
any of the cusps, then the density remains finite until the boundary is 
reached.  For example, for $z=0$ and 
$\rho - a \rightarrow 0_+,~D_2 \simeq \mp b\tau_0 \sqrt{2u(\rho - a)}$ for 
the in and out flows, and hence the density associated with these flows 
increases as ${1\over\sqrt{\rho - a}}$.

Near the cusps, the behaviour depends upon the direction of approach.  
For $z=0$ and $\rho - \rho_0 \rightarrow 0_-,~D_2\simeq 
-2b(\rho_0 -\rho)$ for the up and down flows and hence 
$d \sim {1\over \rho_0 - \rho}$.  For $z=0$ and 
$\rho - \rho_0\rightarrow 0_+$, the density remains finite.  For 
$\rho = \rho_0$ and $z\rightarrow 0_\pm,
~\mid D_2\mid \simeq 3 \left({u^2 b \tau_0^2 s\mid z\mid^2\over 
2}\right)^{1/3}$ for {\it one} of the flows and hence 
$d\sim {1\over\mid z\mid^{2/3}}$.  

The tube caustic collapses to a line caustic in the limit 
$\tau_0\rightarrow 0$ with ${b\over\sqrt{us}}$ fixed.  In this limit, 
$p,q = 0$,  
\begin{equation}
D_2 = -b(u\tau^2 + s\alpha^2) = -2 \sqrt{b^2(\rho -a)^2 + usz^2}~~ ,
\label{4.16}
\end{equation}
and hence the density
\begin{equation}
d(\rho, z) = {1\over 2\pi\rho}~{d^2N\over d\alpha dt_0} 
~{1\over\sqrt{b^2(\rho -a)^2 + usz^2}}~~~\ .
\label{4.17}
\end{equation}
When $p$ and $q$ are finite but much smaller than $a$, Eq.~(4.23) is 
approximately valid for $p,q <<\sigma << a$ since the terms of order 
$\tau_0$ in Eqs.~(\ref{4.5.a}), (\ref{4.6}) and (\ref{4.7}) are small 
in this regime.  

\subsection{Relations between the caustic ring parameters and the initial 
velocity distribution of the dark matter particles}

We saw in section IVA that the flow near the caustic is parametrized by 
the quantities $b,~a,~u,~s$ and $\tau_0$.  In this section we relate these 
quantities to the velocity distribution of the dark matter particles near 
$\alpha = 0$ at the initial time $t_0$.  We assume that the gravitational
potential of the galaxy is spherically symmetric, so that the angular 
momentum of each particle is conserved.  As before, we assume $p,q << a$.  
Thus the particles participating in the flow at the caustic have 
$\mid \alpha \mid  << 1$.  Also the time scale over which the particles 
cross the caustic is short in that limit, which implies that their velocity 
is nearly constant while they do so.

As before, $R(t_0)$ is the turnaround radius of the particles in the 
equatorial plane $(\alpha = 0)$ at time $t_0$.  Thus the radial velocity of 
the particles at $\alpha = 0,~t = t_0$, and $r = R(t_0)$, vanishes.  
Consider the velocity distribution of the particles on the sphere of 
radius $R(t_0)$ at time $t_0$ near $\alpha = 0$:
\begin{eqnarray}
\vec{v} (\theta_0, \varphi_0, t_0) &\equiv& \hat{\varphi}_0
v_\parallel(\alpha)
+ \hat{r}_0 v_r (\alpha) - \hat{\theta}_0 v_\perp (\alpha)\nonumber \\
&=& \hat{\varphi}_0 [v_\parallel (0) + {1\over 2}
v_\parallel^{\prime\prime}(0) 
\alpha^2 + 0 (\alpha^4)] + \hat{r}_0 [{1\over 2} v_r^{\prime\prime} (0) 
\alpha^2 + 0(\alpha^4)] \nonumber\\ 
&+& \hat{\theta}_0 [-\alpha v_\perp^\prime (0) + 0(\alpha^3)]\ ,
\label{4.18}
\end{eqnarray}
assuming axial and reflection symmetry.  $\hat{r}_0,~\hat{\varphi}_0$ and 
$\hat{\theta}_0$ are the unit vectors in spherical coordinates at location 
$(\theta_0,\varphi_0)$ on the sphere.  The dependence of 
$v_\parallel(\alpha), v_\perp(\alpha)$ and $v_r(\alpha)$
upon $t_0$ is not shown explicitly but is understood.  By definition 
$v_\perp^\prime (0) = {dv_\perp\over d\alpha} (0)$, 
$v_\parallel^{\prime\prime} (0) = {d^2 v_\parallel \over d\alpha^2} (0)$,
and so on.  We need only discuss the motion of the particles initially at 
$\varphi_0 = {\pi\over 2}$ since the motion of particles with differing 
values of $\varphi_0$ are trivially related by axial symmetry.  The
particle 
which is initially at 
$\varphi_0 = {\pi\over 2},~\theta_0 = {\pi\over 2} - \alpha$ has 
angular momentum:
\begin{equation}
\vec{\ell} (\alpha) = R~\hat{r}_0 \times \vec{v} ({\pi\over 2} - \alpha, 
{\pi\over 2}, t_0)\equiv \ell (\alpha) \hat{\ell} (\alpha)
\label{4.19}
\end{equation}
where $R \equiv R(t_0)$ and
\begin{equation}
\ell (\alpha) = R\sqrt{v_\parallel^2 (\alpha) + v_\perp^2 (\alpha)}\ ,
\label{4.20}
\end{equation}
and
\begin{equation}
\hat{\ell} (\alpha) = cos \phi (\alpha) (cos \alpha~\hat{z} - sin\alpha~ 
\hat{y}) + sin \phi (\alpha)~\hat{x}\ ,
\label{4.21}
\end{equation}
with
\begin{equation}
\phi (\alpha) = tan^{-1} \left( {v_\perp (\alpha)\over v_\parallel 
(\alpha)}\right)\ .
\label{4.22}
\end{equation}
Let $(x^\prime, y^\prime)$ be Cartesian coordinates in the plane of the 
orbit of that particle, with origin at the galactic center. The
$(x^\prime, y^\prime, z^\prime)$ coordinates are related to the $(x,y,z)$
coordinates by a rotation of angle $\alpha$ about the $\hat{x}$-axis 
followed by a rotation of angle $\phi (\alpha)$ about the 
$\hat{y}^\prime$-axis:
\begin{equation}
\left( \begin{array}{c}
x^\prime \\
y^\prime \\
z^\prime
\end{array}\right) = \left(\begin{array}{ccc}
cos\phi(\alpha)~~ & sin \alpha~sin\phi (\alpha)~~ & -cos \alpha~sin\phi 
(\alpha) \\
0 & cos \alpha & sin \alpha \\
sin\phi (\alpha)~~ & -sin \alpha~cos\phi (\alpha)~~ & cos\alpha~cos\phi 
(\alpha)
\end{array}\right) \left(\begin{array}{c}
x \\
y \\
z \end{array}\right)~~~~ \ .
\label{4.23}
\end{equation}
In the $(x^\prime, y^\prime)$ coordinates, the particle starts at 
$\vec{r}_0 = R\hat{y}^\prime$ with initial velocity $\vec{v}_0 (\alpha) = 
- {\ell(\alpha) \over R} \hat{x}^\prime + v_r (\alpha) \hat{y}^\prime$.  
See Fig. 7.  At the  moment of its closest approach to the galactic
center, the particle moves with approximately constant velocity:
\begin{eqnarray}
x^\prime(\alpha,t_0;t) &=& - r_m~cos\delta~+~V~(t-t_m)~sin\delta 
\nonumber\\
y^\prime(\alpha,t_0;t) &=& - r_m~sin\delta - V~(t-t_m)~cos\delta 
\nonumber\\
z^\prime(\alpha,t_0;t) &=& 0\ ,
\label{4.24}
\end{eqnarray}
where $r_m$ and $t_m$ are the distance and time of closest approach, V is 
the magnitude of velocity then, and $\delta$ is defined in Fig. 7.  The 
quantities $r_m,~t_m,~V$ and $\delta$ depend on $\alpha$ and $t_0$.  On the 
time scale over which the particles cross the ring caustic, the flow is 
very nearly time-independent, i.e. 
$\vec{x} (\alpha, t_0; t) = \vec{x} (\alpha, t-t_0)$.  Hence, we  
replace $t-t_m$ by $t_{0,m} - t_0$ in Eqs. (\ref{4.24}) where 
$t_{0,m}(\alpha)$ is the initial time of the particles which are at their 
closest approach at time $t$.  The flow at fixed $t$ is then given as a 
function of $\alpha$ and $t_0$ by:
\begin{eqnarray}
x^\prime (\alpha, t_0) &=& - r_m (\alpha)~cos\delta(\alpha) +  V(\alpha)~
sin\delta(\alpha)~(t_{0,m}(\alpha) - t_0) \nonumber \\
y^\prime (\alpha, t_0) &=& - r_m (\alpha)~sin\delta(\alpha) -  V(\alpha)~
cos\delta(\alpha)~(t_{0,m} (\alpha) - t_0) \nonumber \\
z^\prime(\alpha,t_0) &=& 0  ~~~~~~~\ .
\label{4.25}
\end{eqnarray}
In Eqs.~(\ref{4.25}) and henceforth, the $t$-dependence of $x^\prime,
y^\prime, r_m, \delta, V$ and $t_{0,m}$ is not explicitly shown.

Using Eqs.~(\ref{4.23}) and (\ref{4.25}), we have:
\begin{eqnarray}
z(\alpha, t_0) &=& r_m (\alpha)[cos\alpha~sin\phi (\alpha)~cos\delta 
(\alpha) - sin\alpha~sin\delta (\alpha)] \nonumber  \\
&+& V (\alpha) (t_0-t_{0,m} (\alpha))
[cos\alpha~sin\phi(\alpha)~sin\delta(\alpha) + 
sin\alpha~cos\delta (\alpha)]\ .
\label{4.26}
\end{eqnarray} 
As in section IVA, we reparametrize the flow 
$t_0 \rightarrow \tau \equiv t_0 - \tau_0(\alpha)$ such that 
$z(\alpha,\tau) = 0$ at $\tau = 0$ for all $\alpha$.  Thus:
\begin{equation}
z(\alpha, \tau) = V (\alpha)~\tau~[cos\alpha~sin\phi(\alpha)~sin\delta 
(\alpha) + sin\alpha~cos\delta (\alpha)]\ .\label{4.27}
\end{equation}
The time shift $\tau_0 (\alpha)$ is given by:
\begin{eqnarray}
&V&(\alpha) \left(\tau_0 (\alpha) - t_{0,m} 
(\alpha)\right)[cos\alpha~sin\phi(\alpha)~sin\delta(\alpha) + 
sin\alpha~cos\delta (\alpha)]\nonumber \\
&+& r_m (\alpha) [cos\alpha~sin\phi(\alpha)~cos\delta(\alpha) - 
sin\alpha~sin\delta (\alpha)] = 0~~~~\ .
\label{4.28}
\end{eqnarray}
Combining Eqs.~(\ref{4.23}), (\ref{4.25}) and (\ref{4.28}), we have:
\begin{eqnarray}
x(\alpha,\tau) = &-& cos\phi~sin\delta~V\tau
- r_m~cos\phi \left[cos\delta - sin\delta ~{cos\alpha~sin\phi~cos\delta 
- sin\alpha~sin\delta\over cos\alpha~sin\phi~sin\delta + 
sin\alpha~cos\delta}\right] \nonumber\\
y(\alpha,\tau) = &-& V\tau (sin\alpha~sin\phi~sin\delta - 
cos\alpha~cos\delta) \nonumber\\
&-& r_m \biggl[sin\alpha~sin\phi~cos\delta + 
cos\alpha~sin\delta \biggr. \nonumber \\
&-& \biggl. (sin\alpha~sin\phi~sin\delta - 
cos\alpha~cos\delta)~{cos\alpha~sin\phi~cos\delta - 
sin\alpha~sin\delta\over cos\alpha~sin\phi~sin\delta + 
sin\alpha~cos\delta} \biggr]
\label{4.29}
\end{eqnarray}
with $V = V(\alpha),~\phi = \phi (\alpha),~r_m = r_m (\alpha)$ and $\delta 
= \delta (\alpha)$.

We now compare Eqs. (\ref{4.27}) and (\ref{4.29}) with Eqs. (\ref{4.5.a}) 
and (\ref{4.6}) to extract $b,~a,~u,~\tau_0$ and $s$.  $\phi (\alpha)$ is an 
odd function of $\alpha$ whereas $r_m (\alpha),~V(\alpha)$ and $\delta 
(\alpha)$ 
are even.  Thus:
\begin{equation}
\phi (\alpha) = \phi^\prime (0) \alpha + 
{1\over 6}~\phi^{\prime\prime\prime} (0) \alpha^3 + 0 (\alpha^5) \ ,
\label{4.30.a} 
\end{equation}
\begin{equation}
r_m (\alpha) = r_m (0) + {1\over 2}~r_m^{\prime\prime} (0) \alpha^2 + 0 
(\alpha^4)\ ,
\label{4.30.b}
\end{equation}
and so on. Comparing Eqs.~(\ref{4.5.a}) and (\ref{4.27}), we have
\begin{equation}
b = V(0) (cos\delta (0) + \phi^\prime(0) sin\delta(0)) \ .
\label{4.31}
\end{equation}
We may rewrite Eqs.~(\ref{4.29}) as
\begin{eqnarray}
x (\alpha, \tau) &=& x_0 (\alpha) + x_1 (\alpha) \tau   \nonumber\\
y(\alpha,\tau) &=& y_0 (\alpha) + y_1 (\alpha) \tau\  ,
\label{4.32}
\end{eqnarray}
with the appropriate definitions of $x_0(\alpha), x_1(\alpha), 
y_0(\alpha)$, and $y_1(\alpha)$.  Therefore
\begin{eqnarray}
\rho (\alpha, \tau)^2 &=& x (\alpha,\tau)^2 + y (\alpha,\tau)^2\nonumber\\
&=& x_0 (\alpha)^2 + y_0 (\alpha)^2 + 
2 \left(x_0 (\alpha) x_1(\alpha) + y_0 (\alpha) y_1 (\alpha)\right)\tau
+ \left(x_1 (\alpha)^2 + y_1 (\alpha)^2\right)\tau^2\ 
,\label{4.33}
\end{eqnarray}
which is to be compared with the square of Eq.~(\ref{4.6}):
\begin{equation}
\rho (\alpha,\tau)^2 = a^2 + ua (\tau-\tau_0)^2 - sa \alpha^2 
+ 0 (\tau^3,~\tau_0^3,~\alpha^2\tau) \ .
\label{4.34}
\end{equation}
For $\alpha =0$ this yields:
\begin{eqnarray}
a^2 + ua\tau_0^2 &=& x_0^2 (0) + y_0^2 (0)\nonumber \\
- ua\tau_0 &=& x_0(0)~x_1 (0) + y_0 (0)~y_1 (0)\nonumber \\
ua &=& x_1^2 (0) + y_1^2 (0)\ .
\label{4.35}
\end{eqnarray}
From Eqs.~(\ref{4.29}) and (\ref{4.32}), we have:
\begin{eqnarray}
x_0 (0) &=& -r_m (0) \left[cos\delta (0) - sin\delta (0)~{\phi^\prime (0)~ 
cos\delta (0) - sin\delta (0)\over \phi^\prime (0)~sin\delta (0) + 
cos\delta (0)}\right]\nonumber \\
x_1 (0) &=& - V(0)~sin\delta (0)\nonumber \\
y_0 (0) &=& - r_m (0) \left[ sin\delta (0) + cos\delta (0) 
~{\phi^\prime(0)~
cos\delta (0) - sin\delta(0)\over \phi^\prime (0)~sin\delta (0) + 
cos\delta (0)}\right]\nonumber \\
y_1 (0) &=& V(0) cos\delta (0)\ .
\label{4.36}
\end{eqnarray}
Hence:
\begin{eqnarray}
a &=& r_m (0) \nonumber\\
u &=& {V(0)^2\over r_m (0)} \nonumber\\
\tau_0 &=& {r_m(0)\over V(0)}~{\phi^\prime (0)~cos\delta (0) - sin\delta 
(0)\over \phi^\prime (0)~sin\delta (0) + cos\delta (0)}~~~\  .
\label{4.37}
\end{eqnarray}
Comparing Eqs.~(\ref{4.33}) and (\ref{4.34}) for $\alpha \neq 0$, we have: 
\begin{equation}
-sa =\left(x_0(0)x_0^{\prime\prime}(0)+y_0(0)y_0^{\prime\prime}(0)\right)\ .
\label{4.38}
\end{equation}
A somewhat lengthy calculation yields:
\begin{eqnarray}
s = &-&~ {r_m^{\prime\prime} (0) (1 + \phi^\prime (0)^2)\over [\phi^\prime 
(0) sin\delta (0) + cos\delta (0)]^2}\nonumber \\
&-& r_m (0)~{\phi^\prime (0) cos\delta (0) - sin\delta (0)\over 
[\phi^\prime (0) sin\delta (0) + cos\delta (0)]^3}~\left\{
- {2\over 3}~\phi^\prime (0) \right. \nonumber\\
&+& \left. {1\over 3} \left( \phi^{\prime\prime\prime} (0) - \phi^\prime 
(0)^3\right) - \left( 1 + \phi^\prime (0)^2\right) \delta^{\prime\prime} 
(0) \right\}\ .
\label{4.39}
\end{eqnarray}
Our derivation of the caustic parameters assumes that $p$ and 
$q = {\sqrt{27}\over 4}~ {b\over\sqrt{us}}~p~~$ are much smaller than $a$.  
Using Eqs.~(\ref{4.37}), we have:
\begin{equation}
p = {1\over 2} a \left({\phi^\prime (0) cos\delta (0) - sin\delta (0)\over
\phi^\prime (0) sin\delta (0) + cos\delta (0)}\right)^2\ .
\label{4.43}
\end{equation}
Thus the treatment requires that $\phi^\prime (0)$ and $\delta (0)$, or at 
least the combination $\phi^\prime (0) cos\delta (0) - sin\delta (0)$, be 
small compared to one.

Eqs.~(\ref{4.31}), (\ref{4.37}) and (\ref{4.39}) express the caustic 
parameters in terms of the values of 
$\phi (\alpha),~r_m (\alpha),~V(\alpha)$ and $\delta (\alpha)$ and their 
first few derivatives at $\alpha =0$.  $\phi (\alpha)$ is given in terms of 
the initial velocity distribution by 
Eq.~(\ref{4.22}).  
Moreover, angular momentum conservation implies:
\begin{equation}
r_m (\alpha) = {\ell (\alpha)\over V(\alpha)}
\label{4.40}
\end{equation}
with $\ell (\alpha)$ given by Eq.~(\ref{4.20}).  Thus, to achieve our goal 
of determining the caustic parameters in terms of the initial velocity 
distribution, it remains to express $V(\alpha)$ and $\delta (\alpha)$ 
in terms of $v_r (\alpha),~ v_\parallel (\alpha)$ and $v_\perp (\alpha)$.

This last step can only be carried out if we adopt a model for the 
galactic gravitational potential $U(r,t)$ in which the particles 
fall.  To illustrate the process, let us adopt the time-independent 
potential:
\begin{equation}
U(r) = v_{rot}^2~\ln \left({R\over r}\right)
\label{4.41}
\end{equation}
which yields (perfectly) flat rotation curves with rotation velocity 
$v_{rot}$.  
Since particle energy is conserved for this potential, we have:
\begin{equation}
V(\alpha) = \left[ 2 v_{rot}^2 \ln \left({R\over a}\right) + v_r^2 
(\alpha) + v_\parallel^2 (\alpha) + v_\perp^2 (\alpha) \right]^{1 \over 
2}\ .
\label{4.42}
\end{equation}
The angle $\delta = \delta \left(\ell,v_r\right)$ was determined 
numerically for the potential $U(r)$ by solving the equations of 
motion of a particle falling from the initial position 
$\vec{r}_0 = R\hat{y}^\prime$ with initial velocity 
$\vec{v}_0 = {-\ell\over R}~\hat{x}^\prime + v_r \hat{y}^\prime$; 
see Fig. 7.  Table I gives $\delta$ as a function of $j = {\ell\over
Rv_{rot}}$ and $\nu = {v_r\over v_{rot}}$.  

A simple velocity distribution is that corresponding to an initially 
rigidly rotating turnaround sphere:  
$v_\perp (\alpha) = 0,~v_r (\alpha) = 0$ and $v_\parallel (\alpha) = 
v_\parallel (0) cos\alpha$.  Let us explore what happens in this 
case.  Since $\phi (\alpha) = 0,$ we have:
\begin{eqnarray}
p &=& {1\over 2} a~tan^2\delta(0)\nonumber\\
q &=& {\sqrt{27}\over 4} ~p 
{cos^2 \delta (0)\over \sqrt{-{r_m^{\prime\prime}(0)\over r_m (0)} 
- tan\delta (0) \delta^{\prime\prime} (0)}}~~~\ .
\label{4.44}
\end{eqnarray}
Typical values in fits of the infall model \cite{sty2,sty,me} to 
observed properties of our galaxy are $j \sim 0.25$ and 
$a \sim {jR \over \sqrt{2 \ln(R/a)}} \sim 0.1 R$.  Table I shows 
that $\delta$ is slowly varying and of order 0.5 then.  For 
$\phi^\prime (0) = 0$, this implies $p \sim 0.15~a$.  For small $j, 
V \simeq v_{rot} \sqrt{2 \ln(R/a)}$ is approximately 
$\alpha$-independent.  For the above special velocity distribution, 
we have then $r_m (\alpha) = r_m (0)~cos\alpha$ and hence 
$q \simeq {\sqrt{27}\over 4}~p~cos^2 \delta (0) \sim p~~~$.

\subsection{Flow some distance away from the caustic}

In this section, we give a qualitative description of the flow 
associated with a caustic ring on distance scales of order $a$, the 
ring radius.  To this effect, we choose the special initial velocity 
distribution corresponding to an initially rigidly rotating turnaround 
sphere:  $v_r (\alpha) = v_\perp (\alpha) = 0, 
~v_\parallel (\alpha) = v_\parallel (0) cos\alpha$.  Also we neglect 
the $\alpha$-dependence of $V$ and set $\delta = 0$.  Table I shows 
that the latter approximation is valid only for $j<<1$, i.e. $a<<R$.
We also take the velocity $\vec{V}$ of each particle to be constant
while it travels distances of order $a$.  These crude approximations 
yield a description of the flow which is topologically correct and 
which can be derived by simple analytical methods, but which is likely 
to agree only qualitatively with actual flows.  In particular, since 
$\phi (\alpha)$ and $\delta (\alpha)$ are set equal to zero, the 
tricusp structure of the caustic is shrunk to a point.  

For the assumed velocity distribution,
\begin{equation}
\ell (\alpha) = \ell_{max} ~cos\alpha
\label{4.46}
\end{equation}
with $\ell_{max} = R v_\parallel (0)$.  Since we neglect the 
$\alpha$-dependence of $V$,
\begin{equation}
r_m (\alpha) = a~cos\alpha\label{4.47}
\end{equation}
with $a = {\ell_{max}\over V}$.  Since we set $\phi (\alpha) = 0$ and 
$\delta (\alpha) = 0$, Eqs.~(\ref{4.27}) and (\ref{4.29}) become:
\begin{eqnarray}
x (\alpha,\tau) &=& - a~cos \alpha \nonumber \\
y (\alpha,\tau) &=& V\tau~cos\alpha\nonumber \\
z (\alpha,\tau) &=& V\tau~sin\alpha\ .
\label{4.48}
\end{eqnarray}
From Eqs.~(\ref{4.48}), one obtains:
\begin{equation}
\rho D_2 (\alpha,\tau) = 
- V~cos\alpha (a^2 sin^2\alpha + V^2\tau^2)
= - V~cos\alpha \sqrt{(r^2-a^2)^2 + 4 a^2 z^2} \ .
\label{4.49}
\end{equation}
Inserting this into Eq.~(\ref{4.1}) and using $d\Omega = 
2\pi~cos\alpha~d\alpha$, we have:
\begin{equation}
d(\rho, z) = {2\over V}~{d^2N\over d\Omega dt_0}~{1\over\sqrt{(r^2-a^2)^2 
+ 4a^2z^2}}\label{4.50}
\end{equation}
where ${d^2N\over d\Omega dt_0}$ is the rate at which the dark matter 
particles fall in per unit solid angle.  

The velocity fields can also be derived.  One finds:
\begin{eqnarray}
v_z &=&  - {\partial z\over\partial\tau} = - V~sin\alpha
= \mp {V\over a} \sqrt{{1 \over 2} \left(a^2 -r^2 + 
\sqrt{(r^2-a^2)^2 + 4a^2 z^2}\right)}\nonumber\\
v_\rho &=& {-1\over 2\rho}~{\partial\rho^2\over\partial\tau} = - 
{V^2\tau~cos^2\alpha\over \rho}\nonumber \\
&=& \mp sign(z) {V\over 2a^2\rho}~\sqrt{{1\over 2} \left(r^2-a^2 + 
\sqrt{(r^2-a^2)^2 + 4a^2z^2}\right)}
\left(r^2 + a^2 - \sqrt{(r^2 -a^2)^2 + 4a^2z^2}\right)\nonumber\\
v_\varphi &=& + \sqrt{V^2 - v_z^2 - v_\rho^2}\ ,
\label{4.51}
\end{eqnarray}
where the $\mp$ signs are for the down and up flows.  In the galactic 
plane 
$(z=0)$ we have:
\begin{eqnarray}
v_z &=& \mp V \sqrt{1 - {r^2\over a^2}}~~~~~~~~~~~~~~~~~~{\rm for}~~~r < 
a\nonumber \\
&=& 0~~~~~~~~~~~~~~~~~~~~~~~~~~~~~~~~~ {\rm for}~~~ r > a \nonumber \\
v_\rho &=& 0 ~~~~~~~~~~~~~~~~~~~~~~~~~~~~~~~~~{\rm for}~~~r < a \nonumber 
\\
&=& \pm V \sqrt{1 - {a^2\over r^2}}~~~~~~~~~~~~~~~~~~{\rm for}~~~r>a 
\nonumber \\
v_\varphi &=& V {r\over a}~~~~~~~~~~~~~~~~~~~~~~~~~~~~~~ {\rm for} ~~~r<a 
\nonumber \\
&=& V {a\over r}~~~~~~~~~~~~~~~~~~~~~~~~~~~~~~{\rm for}~~~r > a\ 
.\label{4.52}
\end{eqnarray}

\section{Gravitational effects of caustics}

The gravitational force per unit mass caused by the particles 
in a zero-velocity dispersion flow $\vec{x}(\vec{\alpha},t)$ is:
\begin{eqnarray}
\vec{F} (\vec{x},t) &=& Gm \int d^3 x^\prime~{d(\vec{x}^\prime,t)\over 
\mid \vec{x}^\prime - \vec{x}\mid^3}~(\vec{x}^\prime - \vec{x})\nonumber\\
&=& Gm \int d^3\alpha~{d^3N\over d\alpha_1 d\alpha_2 
d\alpha_3}~(\vec{\alpha})~{\vec{x} (\vec{\alpha},t) - \vec{x}\over \mid 
\vec{x} (\vec{\alpha},t) - \vec{x} \mid^3}\label{5.1}
\end{eqnarray}
where $m$ is the mass of each particle.  We are particularly interested
in the effect of caustic rings on galactic rotation curves.  Since caustic 
rings migrate only on cosmological time scales \cite{me}, which are much 
longer than gas dynamic time scales, it is reasonable to expect the gas in 
the galactic plane to have relaxed to orbits consistent with the dark matter 
distribution in the caustics. 

If the gas or other material in the galactic plane at radius $r$ moves on a 
circular orbit with velocity $v(r)$, then
\begin{equation}
{v^2(r)\over r} = F(r)
\label{5.2}
\end{equation}
is the inward gravitational force per unit mass at radius $r$.  Let us 
assume that a circular ring caustic lies in the galactic plane at 
radius $a$.  In the spirit of perturbation theory, let 
$v(r) = v_{rot} + v_1 (r)$ and $F(r) = F_0 (r) + F_1 (r)$, where $F_1(r)$ 
is the inward force per unit mass due to the caustic and $v_1(r)$ is the 
perturbation the caustic causes in the rotation curve.  If the matter in 
the caustic were smoothly distributed, the rotation curve would be flat 
with value $v_{rot}$.  At zeroth order, $F_0(r) = {v_{rot}^2\over r}$.  At 
first order,
\begin{equation}
F_1 (r) = {2\over r} ~v_{rot} ~v_1(r)\ .
\label{5.3}
\end{equation}
Let us assume that $p,q, \mid r-a\mid << a$ where $p$ and $q$ are the 
transverse dimensions of the caustic ring.  In that limit we may, when 
calculating $F_1(r)$, neglect the curvature of the ring and pretend 
that it is a straight tube.  After integrating over the length of the 
tube, we have
\begin{equation}
F_1 (r) \simeq 2 Gm \int d\rho\int dz ~{d(\rho, z)\over (r-\rho)^2 + 
z^2}~(r-\rho)\ .\label{5.4}
\end{equation}
Using Eq.~(\ref{4.1}), changing variables 
$(\rho, z) \rightarrow (\alpha, t_0)$, 
neglecting the $(\alpha, t_0)$ dependence of 
${dM\over d\Omega dt_0}={m \over 2\pi cos(\alpha)}~{dN\over d\alpha dt_0}$ 
over the size of the caustic, and approximating $\rho (\alpha, t_0)$ by 
$a$ in the Jacobian factor, we obtain
\begin{equation}
F_1 (r) \simeq {2G\over a}~{dM\over d\Omega dt_0}~\int d\alpha~dt_0~{r- 
\rho (\alpha, t_0)\over (r - \rho (\alpha, t_0))^2 + z^2 (\alpha, t_0)}\ .
\label{5.5}
\end{equation}
Ref. \cite{me} indicates how to extract the prefactor 
${dM\over d\Omega~dt_0}$ from the self-similar infall model.  Here we focus 
on the profile of $v_1(r)$ implied by the structure of caustic rings.  The 
functions $\rho (\alpha, t_0)$ and $z(\alpha, t_0)$ are given by 
Eqs.~(\ref{4.5.a}) and (\ref{4.6}) respectively (replace $\tau$ by $t_0$).  
One finds:
\begin{equation}
v_1 (r) \simeq~{4\pi G\over v_{rot} b}~{dM\over d\Omega~dt_0}~I (\zeta, 
{r-a\over p})\label{5.6}
\end{equation}
where $\zeta = su/b^2$ and
\begin{equation}
I(\zeta,X) \equiv {1\over 2\pi} \int dA~dT~{X-(T-1)^2 + \zeta A^2\over 
[X-(T-1)^2 + \zeta A^2]^2 + 4A^2T^2}\ .\label{5.7}
\end{equation}
Fig. 8 shows $I(\zeta, X)$ as a function of $X$ for $\zeta = 
1.0$.  For $X < 0~(r<a)$ and $X>1~(r>a+p),~I(\zeta, X)$ is constant.  For 
$0<X<1~$(inside the tricusp), $I(\zeta, X)$ rises by an 
amount $\Delta I(\zeta)$.  $\Delta I =1$ for $\zeta = 1$.  Fig. 9 shows 
$\Delta I$ as a function of $\zeta$.  In the limit where the tricusp
collapses to a point $(p\rightarrow 0)$, there is a discontinuity 
in $v_1 (r)$ at the caustic ring radius:
\begin{equation}
\Delta v_1 \simeq ~{4\pi G\over v_{rot} b}~{dM\over d\Omega~dt_0}~\Delta 
I(\zeta)\ .
\label{5.8}
\end{equation}
The profile shown in Fig. 8 should be added to a descending rotation 
curve so that the total rotation curve, with the effect of the caustic 
ring included, remains flat on average.  Fig. 10 gives a qualitative 
description.

\section{Conclusions}

We discussed the appearance of caustics in the flow of collisionless 
particles with negligible velocity dispersion.  Caustics are locations in 
physical space where the density diverges in the limit of zero velocity 
dispersion.  This happens wherever the 3D sheet on which the particles 
lie in 6D phase-space folds back.  The generic caustic is a surface at
the boundary between two regions in physical space, one of which has 
$n$ flows and the other $n+2$ flows.  The density diverges as 
${1\over\sqrt{\sigma}}$, where $\sigma$ is the distance to the surface, 
on the side with $n+2$ flows.  We discussed line caustics as well.  These 
are a degenerate case where the density diverges as $1/\sigma$ where 
$\sigma$ is the distance to the line.  We divided line caustics into two 
types: "attached" ($Im\gamma_\pm = 0$) and "isolated"
($Im\gamma_\pm \neq 0$).  In the latter the density is finite everywhere
except on the line, whereas in the former the line is at the intersection 
of two surface caustics.

We discussed the fall of collisionless dark matter with negligible velocity 
dispersion in and out of a galaxy.  Two types of caustic form:  outer 
caustics which are spherical surfaces surrounding the galaxy and inner 
caustics which are rings.  The caustic rings are located near where the 
particles with the most angular momentum are at their closest approach 
to the galactic center.  The surface of the ring is a closed tube whose 
cross-section is a closed line which has three cusps one of which 
points away from the galactic center.  The tube is the location of a 
generic surface caustic.  Inside the tube there are four flows whereas 
outside there are two flows.  In the limit where the transverse 
dimensions of the tube vanish the ring is a closed line caustic of the 
isolated variety.

We analyzed the ring caustic in detail in the case of axial symmetry about 
the $\hat{z}$-axis, of reflection symmetry $z \rightarrow -z$, and where 
the transverse dimensions, $p$ and $q$, of the tube are much smaller than 
the ring radius $a$.  The caustic is then described by 5 parameters: $a$, 
$u$, $s$, $t_0$ and $b$.  The precise shape of a transverse section of the 
ring in this limit is shown in Fig. 6. The 5 parameters were determined 
in terms of the initial velocity distribution of the infalling dark matter 
particles assuming the gravitational potential of the galaxy to be 
spherically symmetric.  Also, a qualitative description was given of the 
flow of dark matter particles on length scales of order $a$.

Finally we discussed the gravitational effects of caustic rings, in 
particular the perturbation in a galactic rotation curve caused by a
ring lying in the galactic plane.  Figs. 8, 9 and 10 describe 
the shape and size of the bump implied by the tricusp structure of the 
caustic ring.

\section{Acknowledgements}

I thank John Klauder for stimulating comments.  This work was supported in 
part by the US Department of Energy under grant DE-FG02-97ER41029 and by a 
Fellowship from the J. S. Guggenheim Memorial Foundation.

\begin{table}
\caption{Values of $\delta$ as a function of $j$ and $\nu$.}
\vspace{0.5cm}
\begin{tabular}{cccccc}
\tableline
$\nu\rightarrow$ & -0.2 & -0.1 & 0.0 & +0.1 & +0.2 \\
j$\downarrow$ & & & & & \\
0.05 & 0.28 & 0.29 & 0.29 & 0.30 & 0.30 \\
0.10 & 0.34 & 0.36 & 0.37 & 0.38 & 0.38 \\
0.15 & 0.38 & 0.40 & 0.42 & 0.43 & 0.44 \\
0.20 & 0.41 & 0.44 & 0.46 & 0.48 & 0.49 \\
0.25 & 0.43 & 0.46 & 0.49 & 0.52 & 0.54 \\
0.30 & 0.45 & 0.48 & 0.52 & 0.55 & 0.58 \\
0.35 & 0.46 & 0.50 & 0.54 & 0.58 & 0.62 \\
0.40 & 0.46 & 0.51 & 0.56 & 0.61 & 0.65 \\
0.45 & 0.46 & 0.52 & 0.58 & 0.63 & 0.68 \\
0.50 & 0.46 & 0.53 & 0.59 & 0.66 & 0.72 \\
\end{tabular}
\label{tbl1}
\end{table}

\begin{figure}
\vspace{1.5cm}
\epsfxsize=5in
\centerline{\epsfbox{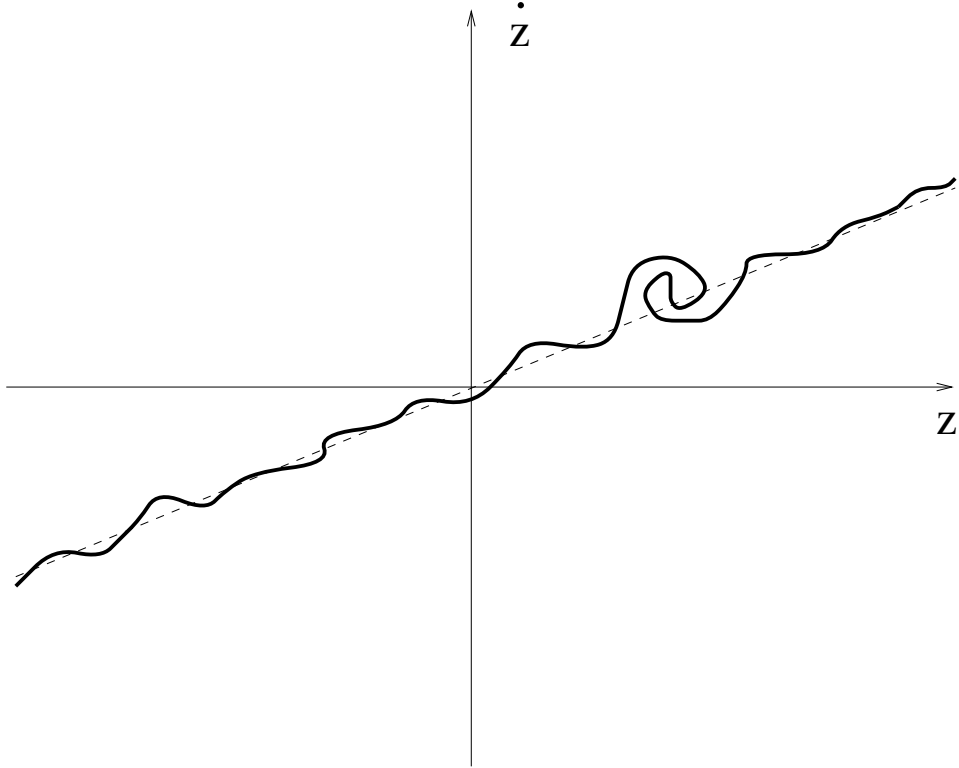}}
\vspace{1.5cm}
\caption{The wiggly line is the intersection of the $(z,\dot{z})$ plane 
with the 3D sheet on which the collisionless dark matter particles lie 
in phase-space.  The thickness of the line is the primordial velocity 
dispersion.  The amplitude of the wiggles in the $\dot{z}$ direction is 
the velocity dispersion associated with density perturbations.  Where an 
overdensity grows in the non-linear regime, the line winds up in clockwise 
fashion.  One such overdensity is shown.}
\label{fig:sheet}
\end{figure}

\begin{figure}
\vspace{1.5cm}
\epsfxsize=5in
\centerline{\epsfbox{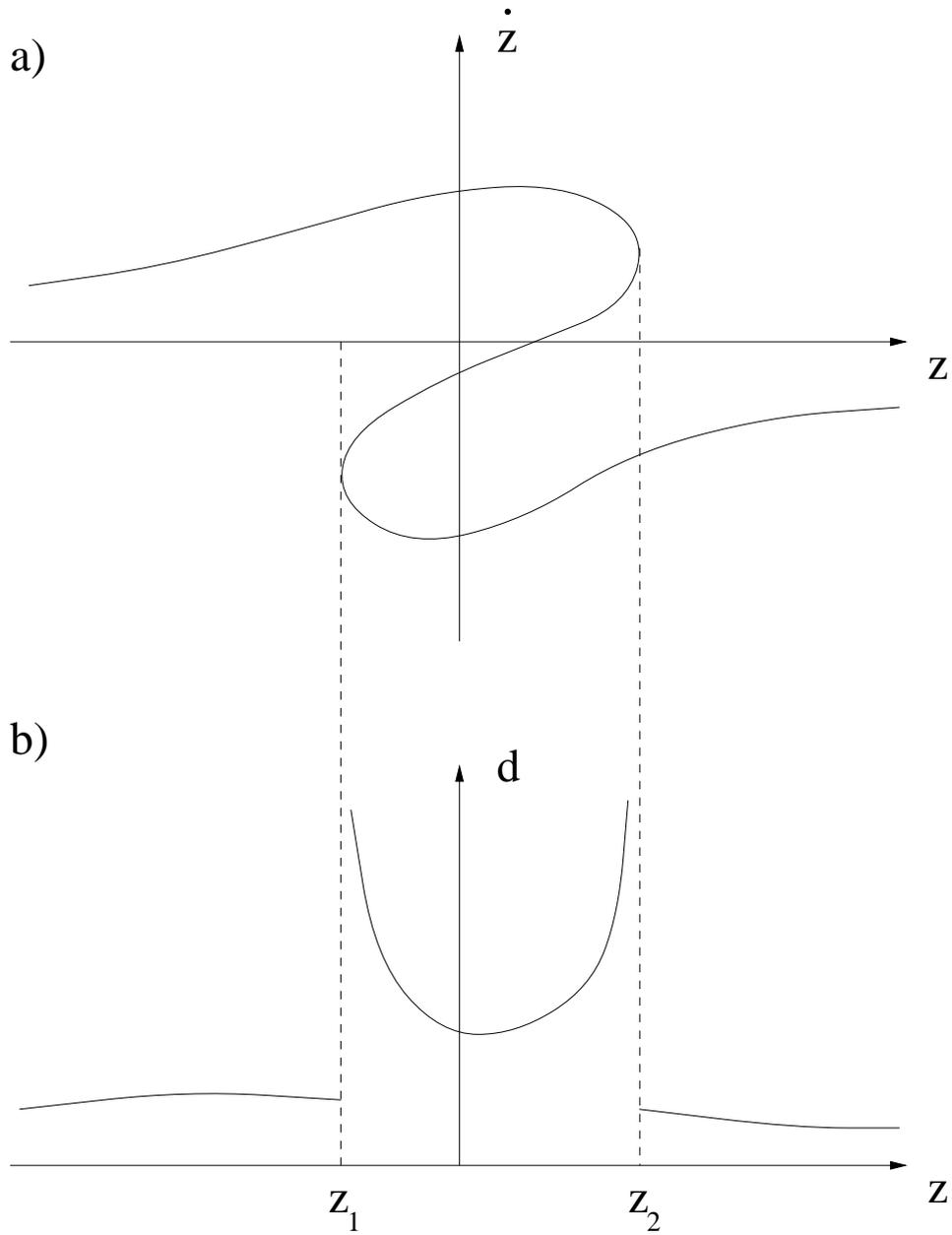}}
\vspace{1.5cm}
\caption{A generic surface caustic in phase-space (a) and in
physical space (b).  The two dimensions ($x$ and $y$) into
which the caustic extends as a surface are not shown.  The
physical space density $d$ diverges at those locations ($z_1$
and $z_2$) where the phase-space sheet folds back.}
\label{fig:Zel}
\end{figure}

\begin{figure}
\vspace{1.5cm}
\epsfxsize=6in
\centerline{\epsfbox{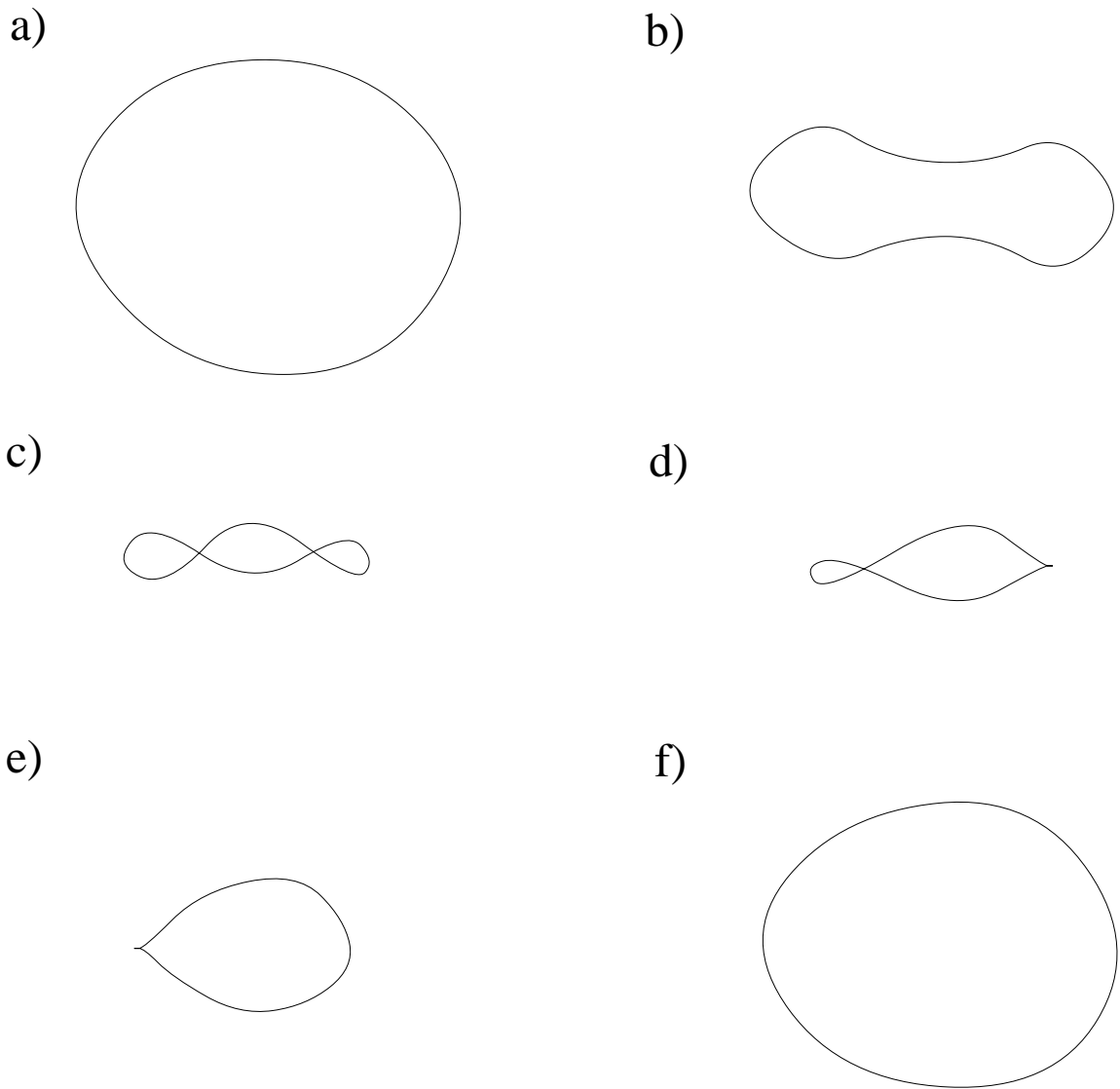}}
\vspace{2.5cm}
\caption{Infall of a turnaround sphere.  The closed lines are at the 
intersections of the sphere with the plane of the figure at six
successive times.  The sphere has net angular momentum about the 
vertical axis.  It crosses itself between frames b) and c).  After 
frame e) the sphere has completed the process of turning itself inside 
out.  The cusps in frames d) and e) are at the intersection of a caustic 
ring with the plane of the figure.}
\label{fig:infall}
\end{figure}

\begin{figure}
\vspace{1.5cm}
\epsfxsize=6in
\centerline{\epsfbox{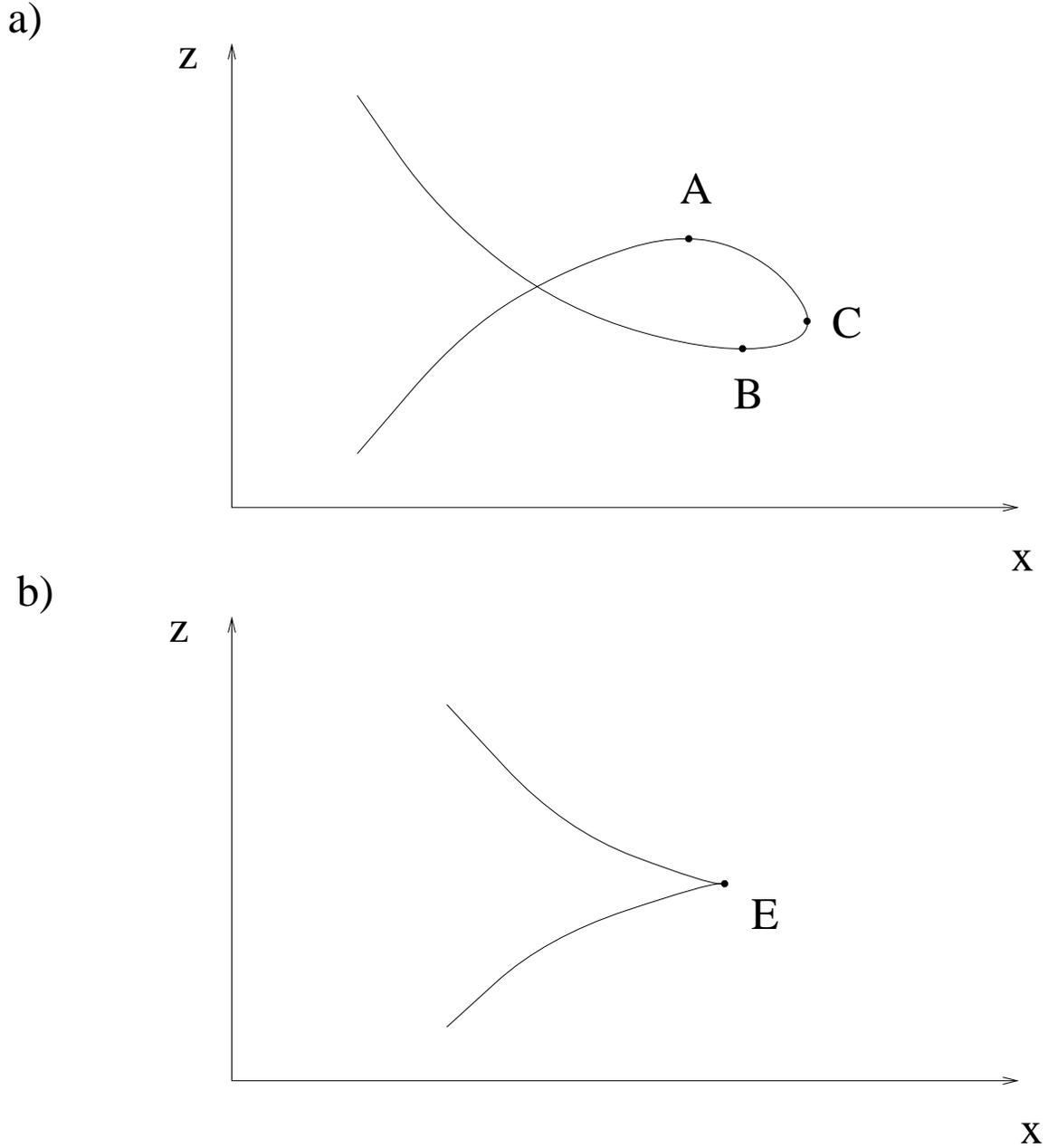}}
\vspace{2.5cm}
\caption{An infall sphere near where and when it completes the process of 
turning itself inside out.  The lines are at the intersections of the 
sphere with the plane of the figure at two successive times, corresponding
to frames c and d in Fig. 3.  $\alpha$ labels points along the line.
${\partial z\over\partial\alpha} = 0$ at points A and B. 
${\partial x\over\partial\alpha} = 0$ at point C.  Points A, B and C move 
to point E, which is thus the location of a caustic since $D=0$ there.}
\label{fig:ABCE}
\end{figure}

\begin{figure}
\vspace{1.5cm}
\epsfxsize=4in
\centerline{\epsfbox{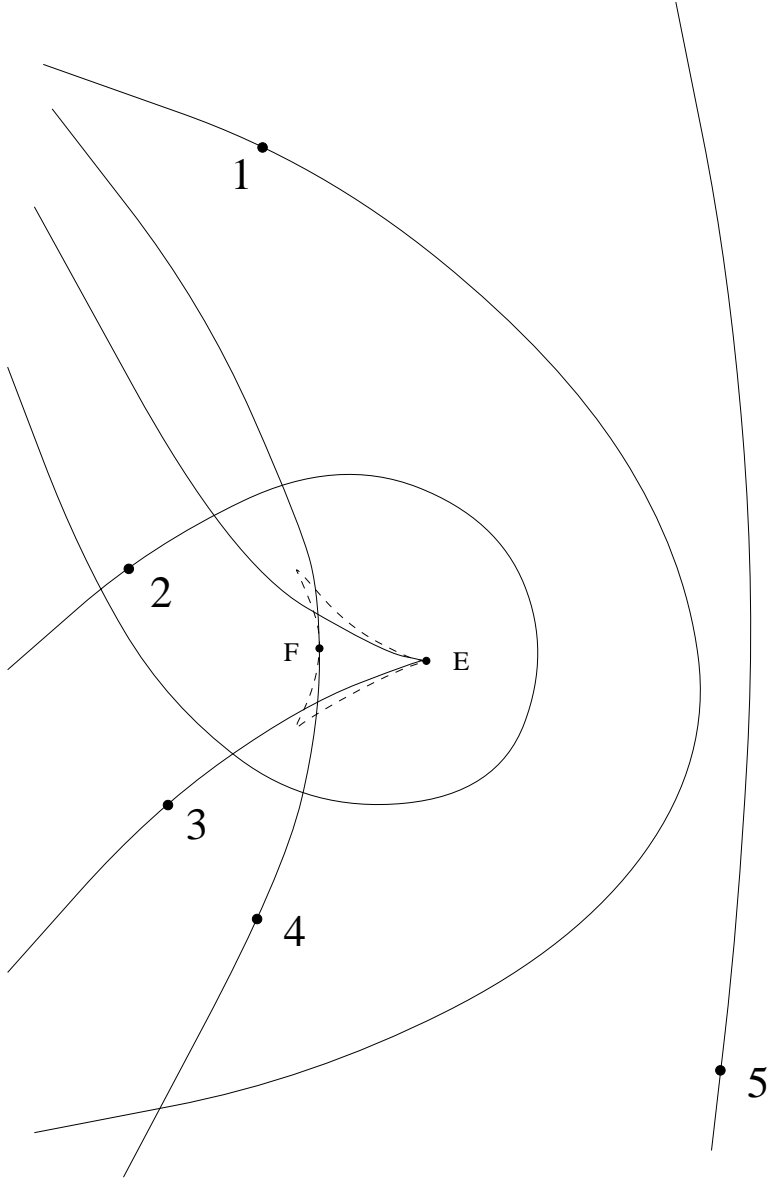}}
\vspace{2.5cm}
\caption{Qualitative description of the flow near a caustic ring.  The
solid lines are at the intersection of five simultaneous infall spheres
corresponding to different initial times $t_{05}<t_{04}< ... <t_{01}$.
The five numbered points are at $\vec{x}(\alpha, t_{0k}), k = 1 ...5$,
for some value of $\alpha$.  Points E and F are defined in the text.  The
closed dashed line is at the intersection of the caustic tube with the
plane of the figure.  There are four flows inside the dashed line whereas
outside there are two.  The galactic center is to the left of the figure.}
\label{fig:flow}

\end{figure}

\begin{figure}  
\vspace{1.5cm}
\psfig{file=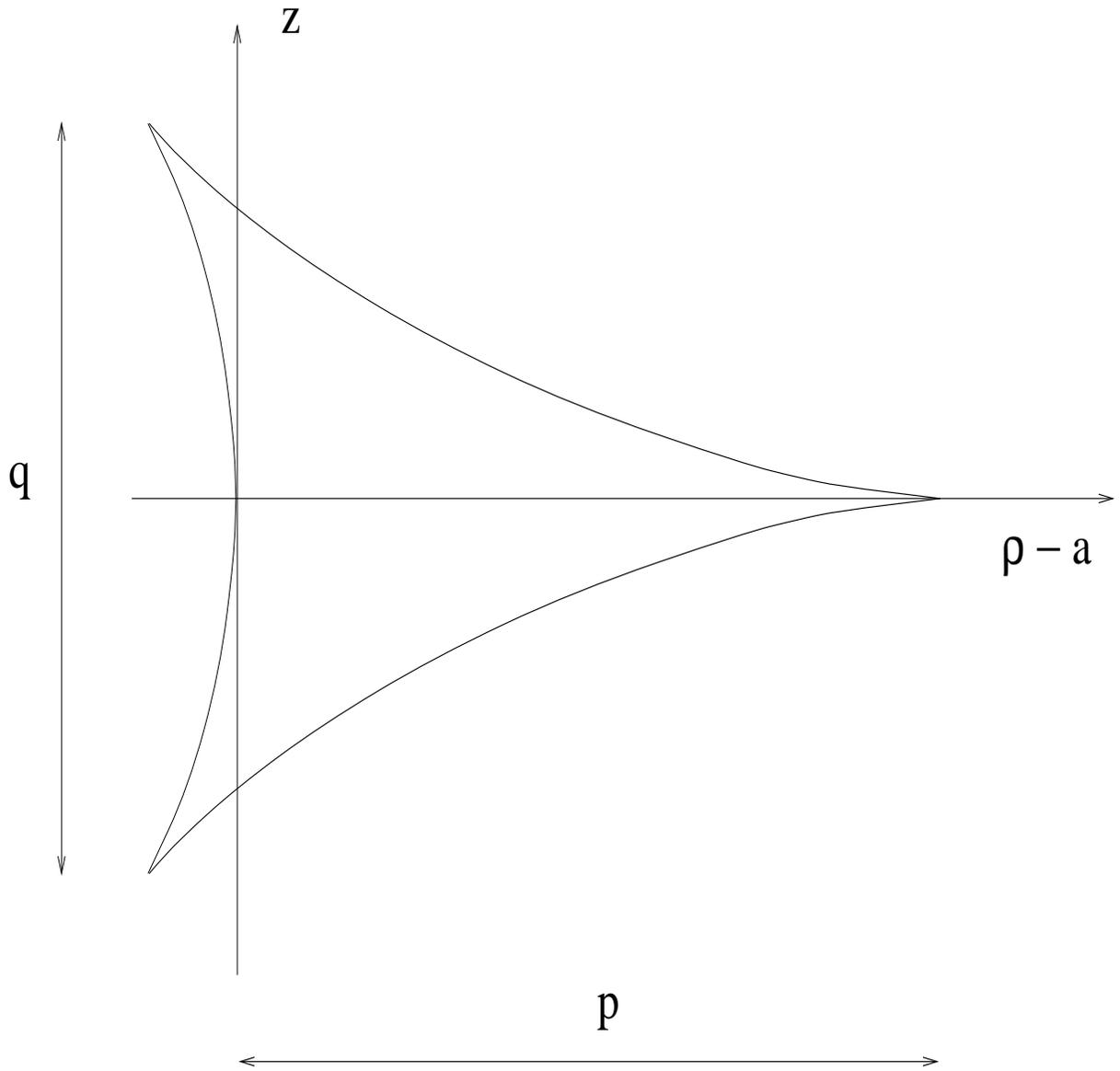,height=6in,width=20in}
%\epsfxsize=10in
%\centerline{\epsfbox{fig6.eps}}
\vspace{2.5cm}
\caption{Cross-section of a caustic ring in the case of axial and
reflection symmetry, and $p,q << a$.}
\label{fig:triscusp}
\end{figure}

\begin{figure}
\vspace{1.5cm}
\epsfxsize=5in
\centerline{\epsfbox{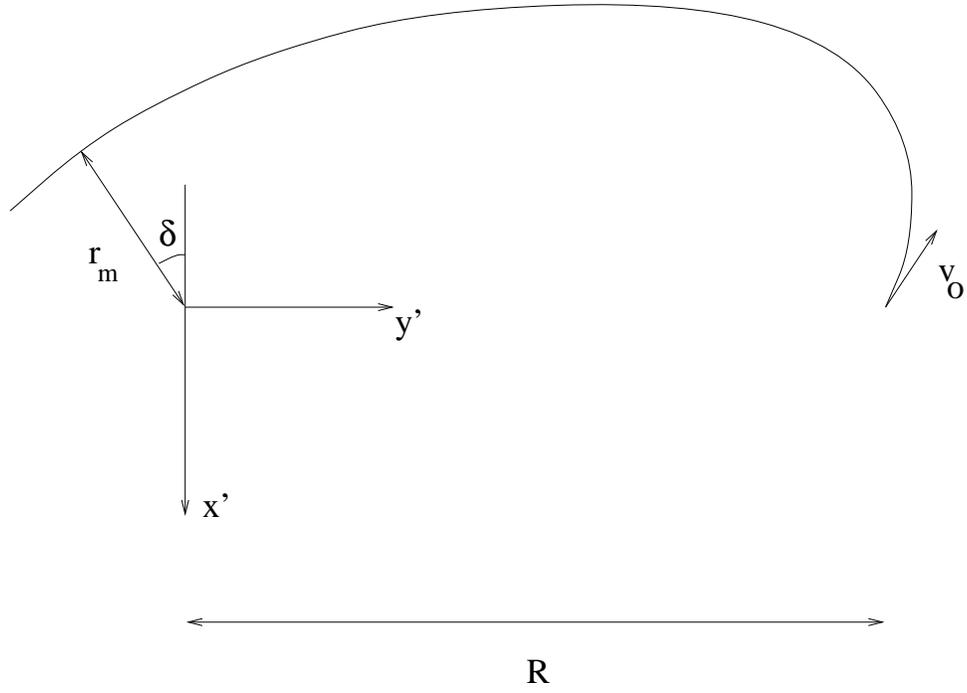}}
\vspace{2.5cm}
\caption{Trajectory of a dark matter particle falling onto a galaxy.
$\vec{v}_0$ is the velocity at the initial time $t_0$.  $r_m$ is the
distance of closest approach to the galactic center.}
\label{fig:delta}
\end{figure}  

\begin{figure}  
\vspace{1.5cm}
\epsfxsize=5in
\centerline{\epsfbox{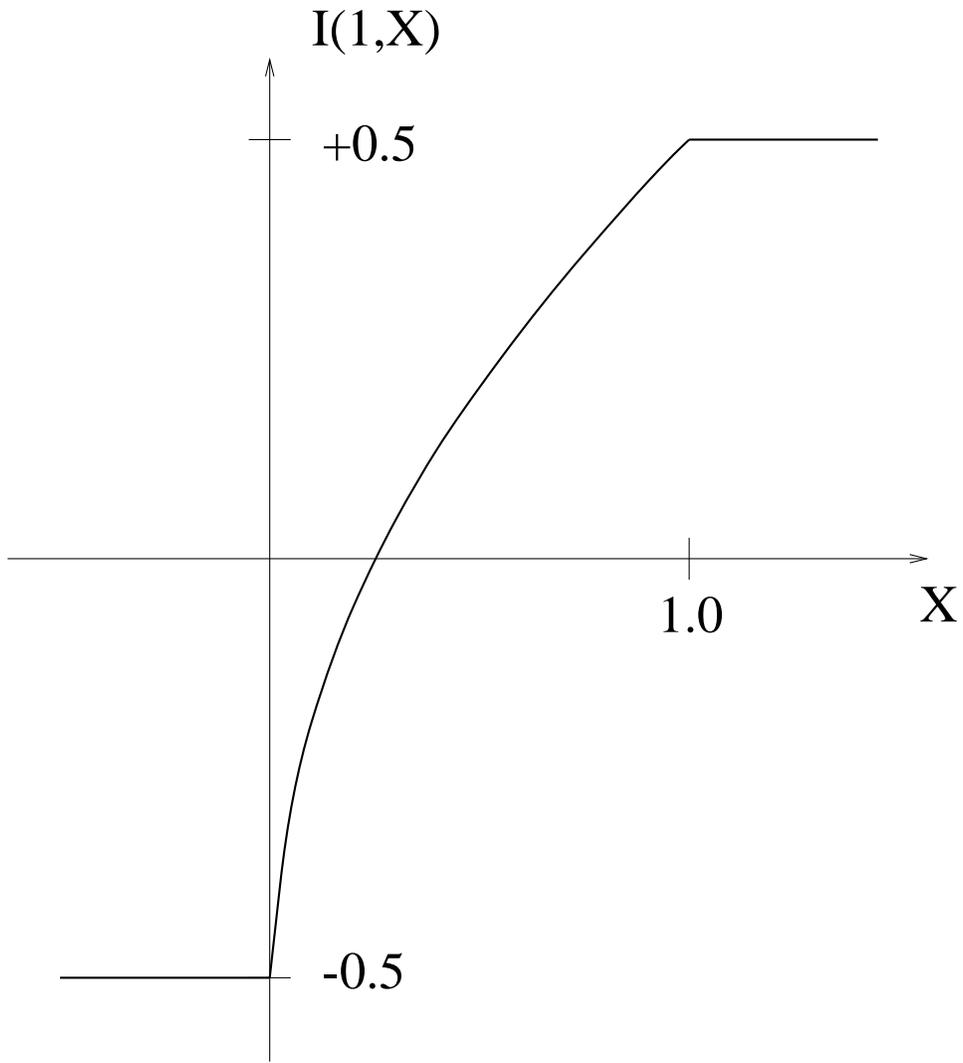}}
\vspace{2.5cm}
\caption{Plot of the function $I(\zeta,X)$ for $\zeta=1.0$.}
\label{fig:Ifunction}
\end{figure}

\begin{figure}
\vspace{1.cm} 
\epsfxsize=3in
\centerline{\epsfbox{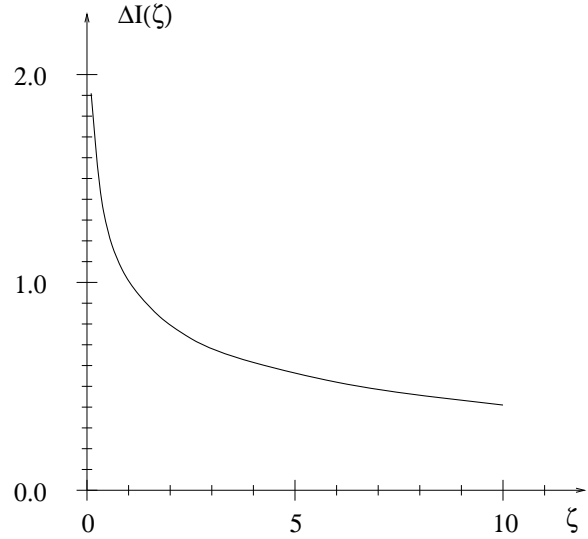}}
\vspace{1.cm}
\caption{Plot of the $\Delta I(\zeta)$ as a function of $\zeta$.}   
\label{fig:deltaI}
\end{figure}

\vspace{2.5cm}

\begin{figure}
\vspace{1.cm} 
\epsfxsize=4in
\centerline{\epsfbox{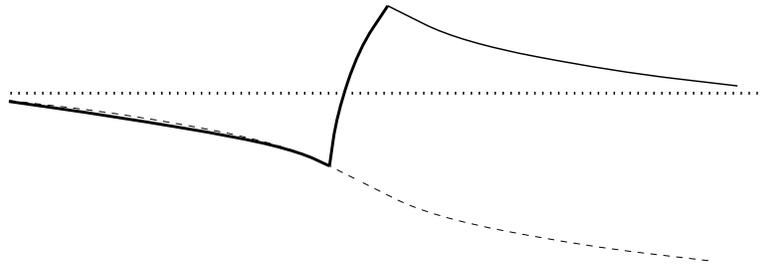}}
\vspace{1.cm}
\caption{Qualitative description of the effect of a caustic ring upon
a galactic rotation curve if the caustic ring lies in the galactic
plane. The horizontal dotted line represents the flat rotation curve if 
the matter in the caustic ring is smoothly distributed.  The descending dashed 
line is the rotation curve after the matter in the caustic ring has been
removed.  The solid line is the rotation curve in the presence of the
caustic ring.  It is the sum of the dashed curve and the profile of Fig.
8.}
\label{fig:rot}
\end{figure}

\end{document}